\documentclass{llncs}
\usepackage[utf8]{inputenc}
\usepackage[T1]{fontenc}

\usepackage{amssymb} 
\usepackage{amsmath}
\usepackage{stmaryrd}
\usepackage{bbm}
\usepackage{graphicx}
\usepackage{paralist}
\usepackage{xcolor}

\usepackage{booktabs}

\usepackage{verbatim} 

\usepackage{tikz}
\usetikzlibrary{arrows,chains,fit}
\usetikzlibrary{automata,positioning}

\usepackage{hyperref}
\usepackage{proof}

\usepackage{fancyvrb}
\usepackage{listings}
\usepackage[linesnumbered,boxed]{algorithm2e}

\usepackage{subcaption}

\usepackage{oz_cav23}
\usepackage{bm}

\usepackage{longtable}

\newcommand{\nonTerm}[1]{\ensuremath{\langle\mathit{#1}\rangle}}

\newcommand{\draftcomment}[3]{{\color{#1}[#2: #3]} 
  \PackageWarning{WARNING: Draft comments visible}{#2: #3}}
\newcommand{\pr}[1]{\marginpar{\draftcomment{blue}{PR}{#1}}}
\newcommand{\dg}[1]{\marginpar{\draftcomment{blue}{DG}{#1}}}
\newcommand{\mdg}[1]{\marginpar{\draftcomment{blue}{DG}{#1}}}
\newcommand{\ze}[1]{\marginpar{\draftcomment{blue}{ZE}{#1}}}
\newcommand{\cl}[1]{\marginpar{\draftcomment{blue}{CL}{#1}}}
\newcommand{\ja}[1]{\marginpar{\draftcomment{blue}{JA}{#1}} }

\iftrue
\renewcommand{\pr}[1]{}
\renewcommand{\dg}[1]{}
\renewcommand{\mdg}[1]{}
\renewcommand{\ze}[1]{}
\renewcommand{\cl}[1]{}
\renewcommand{\ja}[1]{}
\fi

\newcommand{\typeJudg}[2]{#1 : #2}

\newcommand{\instropdef}[1]{$\Omega_{#1} = (G_{\mathit{#1}}, R_{\mathit{#1}}, I_{\mathit{#1}})$}

\newcommand{\eldarica}{\textsc{Eld\-arica}}
\newcommand{\tricera}{\textsc{Tri\-Cera}}
\newcommand{\cpachecker}{\textsc{CPA\-checker}}
\newcommand{\seahorn}{\textsc{Sea\-Horn}}
\newcommand{\jayhorn}{\textsc{Jay\-Horn}}
\newcommand{\rusthorn}{\textsc{Rust\-Horn}}

\newcommand{\monocera}{\textsc{Mono\-Cera}}
\newcommand{\veriabs}{\textsc{Veri\-Abs}}
\newcommand{\korn}{\textsc{Korn}}

\newcommand{\monoc}{\textsc{Mono}}
\newcommand{\tri}{\textsc{Tri}}
\newcommand{\sea}{\textsc{Sea}}
\newcommand{\cpa}{\textsc{CPA}}

\definecolor{mGreen}{rgb}{0,0.6,0}
\definecolor{mGray}{rgb}{0.5,0.5,0.5}
\definecolor{mPurple}{rgb}{0.58,0,0.82}
\definecolor{backgroundColour}{rgb}{0.95,0.95,0.92}
\definecolor{backgroundColourOP}{rgb}{0.97,0.97,0.94}

\lstdefinestyle{CStyle}{
    backgroundcolor=\color{backgroundColour},   
    commentstyle=\color{mGreen},
    keywordstyle=\color{magenta},
    numberstyle=\tiny\color{mGray},
    stringstyle=\color{mPurple},
    basicstyle=\ttfamily,
    breakatwhitespace=false,         
    breaklines=true,                 
    captionpos=b,                    
    keepspaces=true,                 
    numbers=left,                    
    numbersep=5pt,                  
    showspaces=false,                
    showstringspaces=false,
    showtabs=false,                  
    tabsize=2,
    language=C
}

\lstdefinelanguage{ExtWhile}{
    morekeywords={skip, if, else, while, for, assert, assume, const, store, select, aggregate, int},
    sensitive=false,
    morecomment=[l]{//},
    morecomment=[s]{/*}{*/},
}

\lstdefinestyle{ExtWhileStyle}{
    backgroundcolor=\color{backgroundColour},   
    commentstyle=\color{mGreen},
    keywordstyle=\color{magenta},
    numberstyle=\tiny\color{mGray},
    stringstyle=\color{mPurple},
    basicstyle=\ttfamily\footnotesize,
    breakatwhitespace=false,
    breaklines=true,
    captionpos=b,
    keepspaces=true,
    numbers=left,
    numbersep=5pt,
    showspaces=false,
    showstringspaces=false,
    showtabs=false,
    tabsize=2,
    language=ExtWhile
}

\lstdefinestyle{ExtWhileStyleOp}{
    backgroundcolor=\color{backgroundColourOP},   
    commentstyle=\color{mGreen},
    keywordstyle=\color{magenta},
    numberstyle=\tiny\color{mGray},
    stringstyle=\color{mPurple},
    basicstyle=\ttfamily\scriptsize,
    breakatwhitespace=false,
    breaklines=true,
    captionpos=b,
    keepspaces=true,
    numbers=left,
    numbersep=5pt,
    showspaces=false,
    showstringspaces=false,
    showtabs=false,
    tabsize=2,
    language=ExtWhile
}

\definecolor{nodeGray}{rgb}{0.96,0.96,0.96}
\definecolor{cornerGray}{rgb}{0.90,0.90,0.87}
\definecolor{beaublue}{rgb}{0.74, 0.83, 0.9}
\definecolor{mistyrose}{rgb}{1.0, 0.89, 0.88}
\tikzstyle{nodeone} = [rounded corners, text width=1.7cm, minimum height=1.8cm,text centered, draw=black, fill = nodeGray, font = \scriptsize]
\tikzstyle{nodetwo} = [rounded corners, text width=1.5cm, minimum height=2.7cm, minimum width = 5.5cm, text centered, draw=black, fill = beaublue, font = \scriptsize]
\tikzstyle{cornernode} = [rectangle, below right,draw, fill=cornerGray]

\tikzstyle{arrow} = [thick,->,>=stealth]

\newcommand{\assign}{\;\texttt{=}\;}
\newcommand{\eqeq}{\;\texttt{==}\;}
\newcommand{\semic}{\texttt{;}\;}
\newcommand{\add}{\;\texttt{+}\;}
\newcommand{\mult}{\;\texttt{*}\;}

\newif\iftr
\trtrue

\title{Automatic Program Instrumentation for Automatic Verification\iftr\\ (Extended Technical Report)\fi
}

\authorrunning{Program Instrumentation for Automatic Verification}

\author{Jesper Amilon\inst{1},
        Zafer Esen\inst{2},
        Dilian Gurov\inst{1},
        Christian Lidstr\"{o}m\inst{1}, and
        Philipp R\"{u}mmer\inst{2,3}
        }

\institute{KTH Royal Institute of Technology, Stockholm, Sweden \and 
           Uppsala University, Sweden \and University of Regensburg, Germany}

\begin{document}
\maketitle


\vspace*{-2ex}
\begin{abstract}
In deductive verification and software model checking,
dealing with certain specification language constructs can be problematic
when the back-end solver is not 
sufficiently powerful or lacks the required theories. 
One way to deal with this is to transform, for
verification purposes, the program to an equivalent one
not using the problematic constructs, and to reason about 
its correctness instead.
In this paper, we propose instrumentation as a unifying verification paradigm that subsumes various existing ad-hoc approaches, has a clear formal correctness criterion, can be applied automatically, and can transfer back witnesses and counterexamples. 
%
%
We illustrate our approach on the automated verification of programs that involve quantification and aggregation operations over arrays, such as the maximum value or sum of the elements in a given segment of the
array, which are known to be difficult to reason about automatically.
%
%
We implement our approach in the \monocera\ tool, which is tailored to the verification of programs with aggregation, 
and evaluate it on example programs, including SV-COMP programs.
\end{abstract}


\section{Introduction}

\subsubsection{Overview}

Program specifications are often written in expressive, high-level languages: for instance, in temporal logic~\cite{DBLP:reference/mc/2018}, in first-order logic with quantifiers~\cite{DBLP:books/daglib/0022394}, in separation logic~\cite{DBLP:conf/lics/Reynolds02}, or in specification languages that provide extended quantifiers for computing the sum or maximum value of array elements~\cite{DBLP:books/daglib/p/LeavensBR99,acsl}. Specifications commonly also use a rich set of theories; for instance, specifications could be written using full Peano arithmetic, as opposed to bit-vectors or linear arithmetic used in the program. Rich specification languages make it possible to express intended program behaviour in a succinct form, and as a result reduce the likelihood of mistakes being introduced in specifications.

There is a gap, however, between the languages used in specifications and the input languages of automatic verification tools. Software model checkers, in particular, usually require specifications to be expressed using program assertions and Boolean program expressions, and do not directly support any of the more sophisticated language features mentioned. In fact, rich specification languages are challenging to handle in automatic verification, since satisfiability checks can become undecidable (i.e., it is no longer decidable whether assertion failures can occur on a program path), and techniques for inferring program invariants usually focus on simple specifications only.

To bridge this gap, it is common practice to \emph{encode} high-level specifications in the low-level assertion languages understood by the tools. For instance, temporal properties can be translated to B\"uchi automata, and added to programs using ghost variables and assertions~\cite{DBLP:reference/mc/2018}; quantified properties can be replaced with non-determinism, ghost variables, or loops~\cite{DBLP:conf/sas/BjornerMR13,DBLP:conf/sas/MonniauxG16}; sets used to specify the absence of data-races can be represented using non-deterministically initialized variables~\cite{cell2010}. By adding ghost variables and bespoke ghost code to programs~\cite{DBLP:journals/fmsd/FilliatreGP16}, many specifications can be made effectively checkable.


The translation of specifications to assertions or ghost code is today largely designed, or even carried out, by hand. This is an error-prone process, and for complex specifications and programs it is very hard to ensure that the low-level encoding of a specification faithfully models the original high-level properties to be checked. Mistakes have been found even in industrial, very carefully developed specifications~\cite{DBLP:conf/atva/PriyaZSVBG21}\pr{more examples?}, and can result in assertions that are vacuously satisfied by any program. Naturally, the manual translation of specifications also tends to be an ad-hoc process that does not easily generalise to other specifications.

This paper proposes the first general framework to automate the translation of rich program specifications to simpler program assertions, using a process called \emph{instrumentation.} Our approach models the semantics of specific complex operations using program-independent \emph{instrumentation operators,} consisting of (manually designed) rewriting rules that define how the evaluation of the operator can be achieved using simpler program statements and ghost variables. The instrumentation approach is flexible enough to cover a wide range of different operators, including operators that are best handled by weaving their evaluation into the program to be analysed. While instrumentation operators are manually written, their application to programs can be performed in a fully automatic way by means of a search procedure. The soundness of an instrumentation operator is shown formally, once and for all, by providing an \emph{instrumentation invariant} that ensures that the operator can never be used to show correctness of an incorrect program.

\subsubsection{Motivating Example}

\begin{figure}[tb]
\centering
  \begin{minipage}{0.44\linewidth}
\begin{lstlisting}[style=ExtWhileStyle]
// Triangular numbers
i = 0; /*A*/ s = 0; /*B*/
assume(N>0);
while(i < N) {



    i = i + 1; /*C*/

    
    s = s + i;
}


NN = N*N; /*D*/

assert(s == (NN+N)/2);
\end{lstlisting}
  \end{minipage}
  \hfill
  \begin{minipage}{0.50\linewidth}
\begin{lstlisting}[style=ExtWhileStyle]
// Instrumented program
i=0; s=0; x_sq=0; x_shad=0;
assume(N>0);
while(i < N) {
    // Begin-instrumentation
    assert(i == x_shad);
    x_sq   = x_sq + 2*i + 1;
    i      = i + 1;
    x_shad = i;
    // End-instrumentation
    s      = s + i;
}
// Begin-instrumentation
assert(N == x_shad);
NN = x_sq;
// End-instrumentation
assert(s == (NN+N)/2);
\end{lstlisting}
  \end{minipage}
\caption{Program computing triangular numbers, and its instrumented counterpart}
\label{fig:ex-non-linear-arithmetic-code}
\end{figure}




We illustrate our approach on the computation of \emph{triangular numbers}~$s_N = (N^2 + N) / 2$, see left-hand side of \autoref{fig:ex-non-linear-arithmetic-code}. For reasons of presentation, the program has been normalised by representing the square~\lstinline!N*N! using an auxiliary variable~\lstinline!NN!. While mathematically simple, verifying the post-condition~\lstinline!s==(NN+N)/2! in the program turns out to be challenging even for state-of-the-art model checkers, as such tools are usually thrown off course by the non-linear term~\lstinline!N*N!. Computing the value of \lstinline!NN! by adding a loop in line~16 is not sufficient for most tools either, since the program in any case requires a non-linear invariant \verb!0 <= i <= N && 2*s == i*i + i! to be derived for the loop in lines~4--12.

The insight needed to elegantly verify the program is that the value~\lstinline!i*i! can be tracked during the program execution using a ghost variable~\lstinline!x_sq!. For this, the program is instrumented to maintain the relationship~\lstinline!x_sq == i*i!: initially, \lstinline!i == x_sq == 0!, and each time the value of \lstinline!i! is modified, also the variable~\lstinline!x_sq! is updated accordingly. With the value~\lstinline!x_sq == i*i! available, both the loop invariant and the post-condition turn into formulas over linear arithmetic, and program verification becomes largely straightforward. The challenge, of course, is to discover this program transformation automatically, and to guarantee the soundness of the process. For the example, the transformed program is shown on the right-hand side of \autoref{fig:ex-non-linear-arithmetic-code}, and discussed in the next paragraphs.

\begin{figure}[tb]
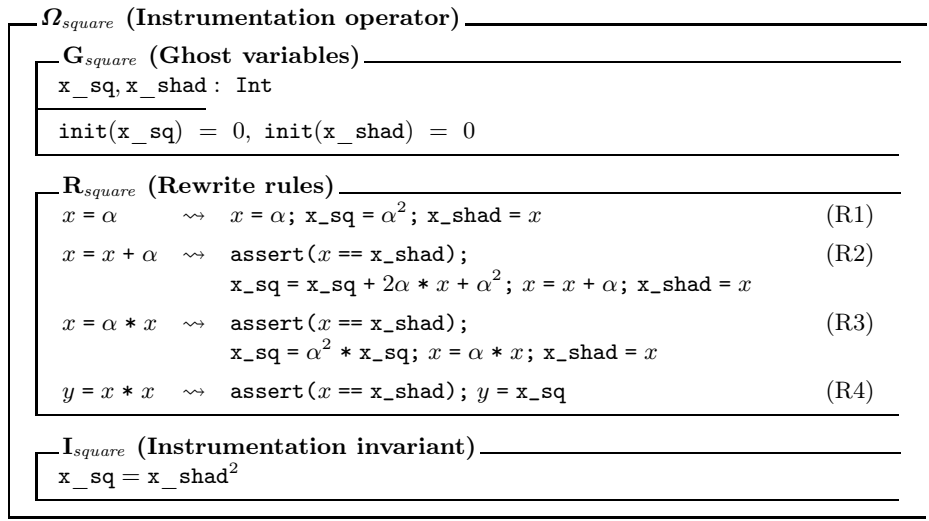

    \centering
\footnotesize
\begin{class}{\bm{\Omega}_{\mathit{square}} \textbf{
(Instrumentation operator)}}
\begin{schema}{\mathbf{G_{\mathit{square}}} \textbf{ (Ghost variables)}}
    \mathtt{x\_sq}, \mathtt{x\_shad:~Int}
    \where
    \mathtt{init(x\_sq)} ~=~ 0,\ 
    \mathtt{init(x\_shad)} ~=~ 0
\end{schema}
\\
\begin{schema}{\mathbf{R_{\mathit{square}}} \textbf{ (Rewrite~rules)}}
\begin{array}{l@{\quad}l@{\quad}l@{\quad\qquad}l}
     x \assign \alpha  &\leadsto& x \assign \alpha\semic \verb!x_sq! \assign \alpha^2\semic \verb!x_shad! \assign x & \text{(R1)} \\[1ex]
     x \assign x \add \alpha  &\leadsto& \texttt{assert(}x \eqeq \verb!x_shad!\texttt{)}\semic  & \text{(R2)}\\
     && \verb!x_sq! \assign \verb!x_sq! \add 2\alpha \mult x \add \alpha^2\semic x \assign x \add \alpha\semic \verb!x_shad! \assign x\\[1ex]
     x \assign \alpha \mult x  &\leadsto& \texttt{assert(}x \eqeq \verb!x_shad!\texttt{)}\semic  & \text{(R3)} \\
      && \verb!x_sq! \assign \alpha^2 \mult \verb!x_sq!\semic x \assign \alpha \mult x\semic \verb!x_shad! \assign x\\[1ex]
     y \assign x\mult x  &\leadsto& \texttt{assert(}x \eqeq \verb!x_shad!\texttt{)}\semic y \assign \verb!x_sq! & \text{(R4)}
\end{array}
\end{schema}
\\
\begin{schema}{\mathbf{I_{\mathit{square}}}\textbf{ (Instrumentation invariant)}}
     \mathtt{x\_sq} = \mathtt{x\_shad}^2
\end{schema}
\end{class}
\caption{Definition of an instrumentation operator $\Omega_{square}$ for tracking squares}
    \label{fig:squareInstrumentation}
    \vspace*{-4ex}
\end{figure}

Our method splits the process of program instrumentation into two parts:
\begin{inparaenum}[(i)]
    \item choosing an \emph{instrumentation operator,} which is defined manually, designed to be program-independent, and induces a space of possible program transformations; and
    \item carrying out an automatic \emph{application strategy} to find, among the possible program transformations, one that enables verification of a program.
\end{inparaenum}

An instrumentation operator for tracking squares is shown in \autoref{fig:squareInstrumentation}, and consists of the declaration of two ghost variables (\verb!x_sq!, \verb!x_shad!) with initial value~$0$, respectively; four rules for rewriting program statements; and the instrumentation invariant witnessing correctness of the operator.  The rewrite rules use formal variables~$x, y$, which can represent arbitrary variables in the program (\lstinline!i!, \lstinline!N!, \lstinline!NN!). An application of the operator to a program will declare the ghost variables in the form of global variables, and then rewrite some chosen set of program statements using the provided rules. Since the statements to be rewritten can be chosen arbitrarily, and since moreover multiple rewrite rules might apply to some statements, rewriting can result in many different variants of a program. In the example, we rewrite the assignments~C, D of the left-hand side program using rewrite rules~(R2) and (R4), respectively, resulting in the instrumented and correct program on the right-hand side.

Instrumentation operators are designed to be \emph{sound,} which means that rewriting a wrong selection of program statements might lead to an instrumented program that cannot be verified, i.e., in which assertions might fail, but instrumentation can never turn an incorrect source program into a correct instrumented program. This opens up the possibility to systematically search for the right program instrumentation. We propose a counterexample-guided algorithm for this purpose, which starts from some arbitrarily chosen instrumentation, checks whether the instrumented program can be verified, and otherwise attempts to fix the instrumentation using a refinement loop. As soon as a verifiable instrumented program has been found, the search can stop and the correctness of the original program has been shown.

The concept of instrumentation invariants is essential for guaranteeing soundness of an operator. Instrumentation invariants are formulas that can (only) refer to the ghost variables introduced by an instrumentation operator, and are formulated in such a way that they hold \emph{in every reachable state of every instrumented program.} To maintain their invariants, instrumentation operators use shadow variables that duplicate the values of program variables. In the operator in \autoref{fig:squareInstrumentation}, the purpose of the shadow variable~\lstinline!x_shad! is to reproduce the value of the program variable whose square is tracked (\lstinline!i!). The rewriting rules introduce guards to detect incorrect instrumentation (the assertions in (R2), (R3), (R4)), which are particular cases in which some update of a relevant variable was missed and not correctly instrumented. The use of shadow variables and guards make instrumentation operators very flexible; in our example, note that instrumentation tracks the square of the value of \lstinline!i! during the loop, but is also used later to simplify the expression~\lstinline!N*N!. This is possible because of the instrumentation invariant and because \lstinline!i == N! holds after termination of the loop, which is verified through the assertion introduced in line~14.

\subsubsection{Contributions and Outline}

The operator shown in \autoref{fig:squareInstrumentation} is simple, and does not apply to all programs, but it can easily be generalised to other arithmetic operators and program statements. The framework presented in this paper provides the foundation for developing a (extendable) library of formally verified instrumentation operators. In the scope of this paper, we focus on two specification constructs that have been identified as particularly challenging in the literature: existential and universal \emph{quantifiers} over arrays, and \emph{aggregation} (or \emph{extended quantifiers}), which includes computing the sum or maximum value of elements in an array.
%
%
%
Our experiments on benchmarks taken from the SV-COMP~\cite{sv-report-22} show that even relatively simple instrumentation operators can significantly extend the capabilities of a software model checker, and often make the automatic verification of otherwise hard specifications easy.

\pr{rewrite contributions}
\dg{Some duplication with first paragraph?}
The contributions of the paper are:
\begin{inparaenum}[(i)]
  \item a general \emph{framework for program instrumentation}, which defines a space of  program transformations that work by rewriting individual statements (\autoref{sec:program-transformation});
  \item an application strategy \emph{search algorithm} in this space, for a given program (\autoref{sec:transformation-strategies});
  \item two \emph{instantiations} of the framework---one for instrumentation operators to handle specifications with \emph{quantifiers} (\autoref{sec:normal-quantifiers}), and one for \emph{extended quantifiers} (\autoref{sec:instrumentation-models});
  \item machine-checked proofs of the correctness of the instrumentation operators for quantifiers~$\forall$ and the extended quantifier~\verb!\max!;
  \item a new \emph{verification tool}, \monocera, that is tailored to the verification of programs with aggregation; and
  \item an \emph{evaluation} of our method and tool on a set of examples, including such from SV-COMP~\cite{sv-report-22} (\autoref{sec:evaluation}).
\end{inparaenum}
\ze{Mention the implementation in \eldarica\ as another contribution?}

\section{Instrumentation Framework}
\label{sec:program-transformation}



The next two sections formally introduce the instrumentation framework. Later, we instantiate the framework for quantification and aggregation over arrays.
We split the instrumentation process into two parts:
\begin{enumerate}
\item
An \emph{instrumentation operator} that defines how to
rewrite program statements
with the purpose of eliminating language constructs that are
difficult to reason about automatically,
but leaves the choice of which occurrences of these statements to
rewrite to the second part (this section).
\item
An \emph{application strategy} for the instrumentation operator, which can be implemented using heuristics or systematic search, among others. The strategy is responsible for selecting the right (if any) program instrumentation from the many possible ones,
\autoref{sec:transformation-strategies} is dedicated to the second part.
\end{enumerate}
Even though instrumentation operators are non-deterministic, we shall guarantee their \emph{soundness:} if the original program has a failing assertion, so will any instrumented program, regardless of the chosen application strategy; that is, instrumentation of an incorrect program will never yield a correct program.
%

We shall also guarantee a
weak form of \emph{completeness}, to the effect that
if an assertion that has not been added to the program by the instrumentation fails in the instrumented program, then it will also fail in the original program. As a result, any counter-example (for such an assertion) produced when verifying the instrumented program can be transformed into a counter-example for the original program.


\begin{table}[tb]
    \caption{Syntax of the core language.}
    \label{tab:programs}
    
    \begin{align*}
        \nonTerm{Type} ~::=~~ & \texttt{Int} \mid \texttt{Bool} \mid \texttt{Array}~\nonTerm{Type}
        \\
        \nonTerm{Expr} ~::=~~ & \nonTerm{DecimalNumber} \mid \texttt{true} \mid \texttt{false} \mid \nonTerm{Variable}
        \\ \mid ~~~ &
        \nonTerm{Expr} \eqeq \nonTerm{Expr} \mid
        \nonTerm{Expr} \;\texttt{<=}\; \nonTerm{Expr} \mid
        \texttt{!}\nonTerm{Expr} \mid \nonTerm{Expr} \;\texttt{\&\&}\; \nonTerm{Expr} 
        \\ \mid ~~~ &
        \nonTerm{Expr} \;\texttt{||}\; \nonTerm{Expr} \mid
        \nonTerm{Expr} \add \nonTerm{Expr} \mid \nonTerm{Expr} \mult \nonTerm{Expr}
        \\ \mid ~~~ &
        \texttt{select(}\nonTerm{Expr}\texttt{,} \nonTerm{Expr}\texttt{)}
        \mid 
        \texttt{store(}\nonTerm{Expr}\texttt{,} \nonTerm{Expr}\texttt{,} \nonTerm{Expr}\texttt{)} 
        \\
        \nonTerm{Prog} ~::=~~ & 
        \texttt{skip} \mid
        \nonTerm{Variable} \assign \nonTerm{Expr}
        \mid
        \nonTerm{Prog}\semic \nonTerm{Prog} \mid
        \texttt{while}~\texttt{(}\nonTerm{Expr}\texttt{)}~\nonTerm{Prog}
        \\ \mid ~~~ &
        \texttt{assert(}\nonTerm{Expr}\texttt{)} \mid
        \texttt{assume(}\nonTerm{Expr}\texttt{)} \mid
        \texttt{if}~\texttt{(}\nonTerm{Expr}\texttt{)}~\nonTerm{Prog}~\texttt{else}~\nonTerm{Prog}
    \end{align*}\vspace*{-10truemm}
\end{table}


\subsection{The Core Language}
\label{subsec:programming-language}

While our implementation works on programs represented as constrained Horn clauses~\cite{DBLP:conf/birthday/BjornerGMR15}, i.e., is language-agnostic,
for readability purposes we present our approach in the setting of an imperative core programming language with data-types for unbounded integers, Booleans, and arrays, and \texttt{assert} and \texttt{assume} statements. The language is deliberately kept simple, but is still close to
standard C. The main exception is the semantics of arrays: they are defined here to be 
\emph{functional} and therefore represent a value type. Arrays have integers as index type
and are unbounded, and their signature and semantics are otherwise borrowed from the SMT-LIB theory of extensional arrays~\cite{smtlib}:
\begin{itemize}
\item
\makebox[0.7\linewidth][l]{%
  \emph{Reading} the value of an array~\texttt{a} at index \texttt{i}:} \texttt{select(a, i)};
\item
\makebox[0.7\linewidth][l]{%
\emph{Updating} an array~\texttt{a} at index~\texttt{i} with 
a new value~\texttt{x}:} \texttt{store(a, i, x)}.
\end{itemize}

The complete syntax of the core language is given in \autoref{tab:programs}.
Programs are written using a vocabulary~$\mathcal{X}$ of typed program variables;
the typing rules of the language are given in
\iftr Appendix~\ref{app:typing-rules}\else \cite{techreport}\fi.
As syntactic sugar, we sometimes write \verb!a[i]! instead of \texttt{select(a, i)}, and 
\verb!a[i] = x! instead of \texttt{a = store(a, i, x)}. 

We denote by $D_{\sigma}$ the domain of a program type~$\sigma$. The domain
of an array type~$\mathtt{Array}~\sigma$ is the set of
functions~$f : \mathbbm{Z} \to D_{\sigma}$.

\paragraph{Semantics.}
%
We assume the Flanagan-Saxe \emph{extended execution model} of programs with 
\texttt{assume} and 
\texttt{assert} statements (see, e.g., \cite{fla-sax-01-popl}),  in which
%
executing 
an \texttt{assert} 
statement with an argument that evaluates to \textsf{false} 
\emph{fails}, i.e., terminates abnormally. An \texttt{assume}
statement with an argument that evaluates to \textsf{false} has the same
semantics as a non-terminating loop.
Partial correctness properties of programs are expressed using \emph{Hoare triples} $\{\mathit{Pre}\} \;P\; \{\mathit{Post}\}$, which state that an execution of $P$, starting in a state satisfying $\mathit{Pre}$, never fails, and may only terminate in states that satisfy $\mathit{Post}$.
As usual, a program~$P$ is considered \emph{(partially) correct} if the Hoare triple~$\{\mathit{true}\} \;P\; \{\mathit{true}\}$ holds.


The evaluation of program expressions is modelled
using a function~$\llbracket \cdot \rrbracket_s$ that maps program
expressions~$t$ of type~$\sigma$ to their value~$\llbracket t \rrbracket_s \in D_{\sigma}$
in the state~$s$.
%
%


\subsection{Instrumentation Operators}
\label{sec:instrumentation_operators}
An instrumentation operator defines schemes to rewrite programs
while preserving the meaning of the existing program assertions. Without
loss of generality, we restrict program rewriting to assignment
statements. Instrumentation can introduce
\emph{ghost state} by adding arbitrary fresh variables to the program.
%
The main part of an instrumentation consists of \emph{rewrite rules},
which are schematic rules $r \;\texttt{=}\; t \leadsto s$, where
the meta-variable~$r$ ranges over program variables, $t$~is an expression
that can contain further meta-variables,
and $s$ is a schematic program in which
the meta-variables from $r \;\texttt{=}\; t$ might occur. Any assignment that matches $r \;\texttt{=}\; t$
can be rewritten to~$s$.


\begin{definition}[Instrumentation Operator]
\label{def:instr-op}
An \emph{instrumentation operator}
is a tuple
%
\instropdef{}, where:
\begin{itemize}
    \item[($i$)]
    $G = \langle (\mathtt{x}_1, \mathit{init}_1), \ldots, (\mathtt{x}_k, \mathit{init}_k) \rangle$
    is a tuple of pairs of ghost variables and their initial values;
    \item[($ii$)]
    $R$
    is a set of rewrite rules
    $r \;\texttt{=}\; t \leadsto s$, where $s$ is a program operating on the ghost
    variables~$\mathtt{x}_1, \ldots, \mathtt{x}_k$ (and containing meta-variables from $r \;\texttt{=}\; t$);
    \item[($iii$)]
    $I$ is a formula over the ghost variables~$\mathtt{x}_1, \ldots, \mathtt{x}_k$, called the
    \emph{instrumentation invariant.}
    
    \end{itemize}
The rewrite rules~$R$ and the invariant~$I$ must adhere to the following constraints:
\begin{enumerate}
    \item The instrumentation invariant $I$ is satisfied by the initial ghost values, i.e., it holds in the state
     $\{\mathtt{x}_1 \mapsto \mathit{init}_1, \ldots, \mathtt{x}_k \mapsto \mathit{init}_k\}$.
    \item For all rewrites
        $r \assign t \leadsto s \in R$
        the following hold:
        \begin{enumerate}
            \item
                $s$ terminates (normally or abnormally) for 
                pre-states satisfying $I$, assuming that all meta-variables
                are ordinary program variables.
            \item $s$ does not assign to variables other than
                $r$ or the ghost variables~$\mathtt{x}_1, \ldots, \mathtt{x}_k$.
            \item $s$ preserves the instrumentation invariant:
                $\{ I \}\ s'\ \{ I \}$, where $s'$ is $s$
                with every $\texttt{assert(}e\texttt{)}$ statement replaced by an $\texttt{assume(}e\texttt{)}$ statement.
            \item $s$ preserves the semantics of the assignment~$r \assign t$: the Hoare triple
                $\{ I \} \;\texttt{z} \assign t\semic s' \; \{ \texttt{z} = r \}$,
                where $\texttt{z}$ is a fresh variable, holds.
        \end{enumerate}
\end{enumerate}
\end{definition}

The conditions imposed in the definition ensure
that all instrumentations are \emph{correct}, in the sense
that they are sound and weakly complete, as we show below. 
In particular, the instrumentation invariant guarantees that the rewrites of program
statements are \emph{semantics-preserving} w.r.t. the original
program, and thus, the execution of any \lstinline{assert}
statement of the original program has the same effect
before and after instrumentation.
%
Observe that the conditions can themselves be deductively verified to hold
for each concrete instrumentation operator, and that this check is \emph{independent} of the programs to be instrumented, so that
an instrumentation operator can be proven correct once and for all.

An instrumentation operator~$\Omega$ does itself not define 
which occurrences of program statements are to be rewritten, 
but only how they are rewritten.
Given a program~$P$ and the operator~$\Omega$, an instrumented program~$P'$
is derived by carrying out the following two steps:
\begin{inparaenum}[(i)]
\item variables $\mathtt{x}_1, \ldots, \mathtt{x}_k$ and the 
 assignments $\mathtt{x}_1 \assign \mathit{init}_1\semic \ldots\semic \mathtt{x}_k \assign \mathit{init}_k$
  are added at the beginning of the program, and
\item some of the assignments in~$P$, to which a rewriting rule~$r \assign t \leadsto s$ in $\Omega$
is applicable, are replaced by~$s$, substituting
meta-variables with the actual terms occurring in the assignment.
\end{inparaenum}
We denote by~$\Omega(P)$ the set of all instrumented programs~$P'$ that can be 
derived in this way.
An example of an instrumentation operator and its
application was shown \autoref{fig:ex-non-linear-arithmetic-code}
and \autoref{fig:squareInstrumentation}.

\subsection{Instrumentation Correctness}
\label{subsec:instrumentation-soundness}

Verification of an instrumented program produces one of two
possible results: a \emph{witness} if verification is successful, or a \emph{counter-example} otherwise.
A witness consists of the inductive invariants needed to verify the
program, and is presented in the context of the programming language:
it is translated back from the back-end theory used by the
verification tool,
and is a formula over the program variables and the ghost variables
added during instrumentation.
A counter-example is an execution trace leading to a failing assertion.

\begin{definition}[Soundness]
\label{def:instr-op-correctness}
An instrumentation operator~$\Omega$ is called \emph{sound} 
if for every program~$P$ and instrumented program~$P' \in \Omega (P)$,
whenever there is an execution of~$P$ where some \texttt{assert} 
statement fails, then 
there also is an execution of~$P'$ where some 
\texttt{assert} statement fails.
\end{definition}

Equivalently, existence of a witness for an instrumented program
entails existence of a witness for the original program,
in the form of a set of inductive invariants solely over the program variables.
Notably, because of the semantics-preserving nature
of the rewrites under the instrumentation invariant, a witness for the
original program can be derived from one for the instrumented
program. 
One such back-translation is to add the instrumentation invariant
as a conjunct to the original witness, and to existentially
quantify over the ghost variables.
%

\paragraph{Example.}
To illustrate the back-translation,
we return to the instrumentation operator from
\autoref{fig:squareInstrumentation} and the example
program from \autoref{fig:ex-non-linear-arithmetic-code}.
The witness produced by our verification tool in this case
is the formula:
\begin{equation*}
\begin{split}
\mathtt{i} = \mathtt{x\_shad} \wedge
\mathtt{x\_sq} + \mathtt{x\_shad} = 2s\wedge
\mathtt{N} \geq \mathtt{i} \wedge
\mathtt{N} \geq \mathtt{1} \wedge
2\mathtt{s} \geq \mathtt{i} \wedge
\mathtt{i} \geq  0
\end{split}
\end{equation*}
After conjoining the instrumentation invariant
$\mathtt{x\_sq} = \mathtt{x\_shad}^2$
and existentially quantifying over the involved ghost variables, 
we obtain
an inductive invariant that is sufficient to verify the original program:
\begin{equation*}
\begin{split}
\exists x_\mathrm{sq}, x_\mathrm{shad}.\; (
\mathtt{i} = x_\mathrm{shad} \wedge
x_\mathrm{sq} + x_\mathrm{shad} = 2s \wedge~~~~~~~~~~
\\
\mathtt{N} \geq \mathtt{i} \wedge
\mathtt{N} \geq \mathtt{1} \wedge
2\mathtt{s} \geq \mathtt{i} \wedge
\mathtt{i} \geq  0
\wedge x_\mathrm{sq} = x_\mathrm{shad}^2)
\end{split}
\end{equation*}
%
\begin{definition}[Weak Completeness]
\label{def:weak-completeness}
The operator~$\Omega$ is called \emph{weakly complete} 
if for every program~$P$ and instrumented program~$P' \in \Omega(P)$,
whenever an \texttt{assert} statement that
has not been added to the program by the instrumentation fails
in the instrumented
program~$P'$, then it also fails in the original program~$P$.
\end{definition}
Similarly to the back-translation of invariants,
when verification fails, counter-examples
for assertions of the original program, found during
verification of the instrumented program, can be translated back to
counter-examples for the original program.
We thus obtain the following result.
%
\begin{theorem}[Soundness and weak completeness]
\label{thm:soundness-weak-completness}
Every instrumentation operator~$\Omega$ is sound and weakly complete.
\end{theorem}
\paragraph{Proof.}
Let~\instropdef{} be an instrumentation operator.
Since $I$ is a formula over ghost variables only,
which holds initially and is preserved by all rewrites,
$I$~is an invariant of the fully instrumented program.
This entails that rewrites of assignments are semantics-preserving.
Furthermore, since instrumentation code only assigns to ghost variables
or to $r$ (i.e., the left-hand side of the original
statement), program variables have the same values
in the instrumented program as in the original one.
Furthermore, since all rewrites are terminating under~$I$, the instrumented program will terminate if and only if the original
program does.

In the case when verification succeeds, and a witness
is produced, weak completeness follows vacuously.
A witness consists of the inductive invariants sufficient to verify
the instrumented program.
Thus, they are also sufficient to verify the
the assertions existing in the original program,
since assertions are not rewritten and all program variables
have the same valuation in the original and the instrumented programs.
%
Since a witness for the instrumented program can be back-translated
to a witness for the original program, any failing assertion
in the original program must also fail in the instrumented program,
and $\Omega$ is therefore sound.

In the case when verification fails, soundness follows vacuously,
and if the failing assertion was added during instrumentation, also
weak completeness follows.
If the assertion existed in the original program, since such assertions
are not rewritten, and since program variables have the same values
in the instrumented program as in the original program,
then any counter-example for the instrumented program is also
a counter-example for the original program, when projected
onto the program variables.
\hfill \qed


\section{Instrumentation Application Strategies}
\label{sec:transformation-strategies}



\begin{figure}[tb]
\begin{algorithm}[H]
  \SetKwInOut{Input}{Input}
  
  \Input{Program~$P$; statements~$S$; instrumentation space~$R$; oracle~$\mathit{IsCorrect}$.}
  \KwResult{Instrumentation~$r \in R$ with $\mathit{IsCorrect}(P_r)$; $\mathit{Incorrect}$; or $\mathit{Inconclusive}$.}
  
  \Begin{
  $\mathit{Cand} \leftarrow R$\;
  \While{$\mathit{Cand} \not= \emptyset$}{
    pick $r \in \mathit{Cand}$\;
    \uIf{$\mathit{IsCorrect}(P_r)$}{
        \Return{$r$}\;
    }
    \Else{
        $\mathit{cex} \leftarrow $ counterexample path for $P_r$\;
        \uIf{failing assertion in $\mathit{cex}$ also exists in $P$}{
            \tcc{$\mathit{cex}$ is also a counterexample for $P$}
            \Return{$\mathit{Incorrect}$}\;
        }
        \Else{
            \tcc{instrumentation on $\mathit{cex}$ may have been incorrect}
            $C' \leftarrow \{ p \in C \mid \mathit{ins}_r(p) \text{~occurs on~} \mathit{cex} \}$\;
            $\mathit{Cand} \leftarrow \mathit{Cand} \setminus \{ r' \in \mathit{Cand} \mid r(s) = r'(s) \text{~for all~} p \in C' \}$\;
        }
    }
  }
  \Return{$\mathit{Inconclusive}$}\;
  } 
  
  \caption{Counterexample-guided instrumentation search}
  \label{alg:instrumentation_search}
\end{algorithm}
\vspace{-4ex}
\end{figure}

We will now define a counterexample-guided search procedure to discover applications of instrumentation operators that make it possible to verify a program.

For our algorithm, we assume that we are given an oracle~$\mathit{IsCorrect}$ that is able to check the correctness of programs after instrumentation. Such an oracle could be approximated, for instance, using a software model checker. The oracle is free to ignore the complex functions we are trying to eliminate by instrumentation; for instance, in \autoref{fig:ex-non-linear-arithmetic-code}, the oracle can over-approximate the term~\lstinline!N*N! by assuming that it can have any value. We further assume that $C$ is the set of control points of a program~$P$ corresponding to the statements to which a given set of instrumentation operators can be applied. For each control point~$p \in C$, let $Q(p)$ be the set of rewrite rules applicable to the statement at $p$, including also a distinguished value~$\bot$ that expresses that $p$ is not modified. For the program in \autoref{fig:ex-non-linear-arithmetic-code}, for instance, the choices could be defined by $Q(\mathtt{A}) = Q(\mathtt{B}) = \{ \mathrm{(R1)} , \bot \}$, $Q(\mathtt{C}) = \{ \mathrm{(R2)} , \bot \}$, and $Q(\mathtt{D}) = \{ \mathrm{(R4)} , \bot \}$, referring to the rules in \autoref{fig:squareInstrumentation}. Any function~$r : C \to \bigcup_{p \in C} Q(p)$ with $r(p) \in Q(p)$ will then define one possible program instrumentation. We will denote the set of well-typed functions~$C \to \bigcup_{p \in C} Q(p)$ by $R$, and the program obtained by rewriting~$P$ according to $r \in R$ by $P_r$. We further denote the control point in $P_r$ corresponding to some $p \in C$ in $P$ by $\mathit{ins}_r (p)$.

\autoref{alg:instrumentation_search} presents our algorithm to search for instrumentations that are sufficient to verify a program~$P$. The algorithm maintains a set~$\mathit{Cand} \subseteq R$ of remaining ways to instrument~$P$, and in each loop considers one of the remaining elements~$r \in \mathit{Cand}$ (line~4). If the oracle manages to verify $P_r$ in line~5, due to soundness of instrumentation the correctness of $P$ has been shown (line~6); if $P_r$ is incorrect, there has to be a counterexample ending with a failing assertion (line~8). There are two possible causes of assertion failures: if the failing assertion in $P_r$ already existed in $P$, then due to the weak completeness of instrumentation also $P$ has to be incorrect (line~10). Otherwise, the program instrumentation has to be refined, and for this from $\mathit{Cand}$ we remove all instrumentations~$r'$ that agree with $r$ regarding the instrumentation of the statements occurring in the counterexample (line~13).

Since $R$ is finite, and at least one element of $\mathit{Cand}$ is eliminated in each iteration, the refinement loop terminates. The set~$\mathit{Cand}$ can be exponentially big, however, and therefore should be represented symbolically (using BDDs, or using an SMT solver managing the set of blocking constraints from line~13).

We can observe soundness and completeness of the algorithm w.r.t.\ the considered instrumentation operators (proof in \iftr Appendix~\ref{app:instrSearchCorrectness}\else \cite{techreport}\fi):
\begin{lemma}[Correctness]
    \label{lem:searchCorrectness}
    If \autoref{alg:instrumentation_search} returns an instrumentation~$r \in R$, then $P_r$ and $P$ are correct. If \autoref{alg:instrumentation_search} returns $\mathit{Incorrect}$, then $P$ is incorrect.
    If there is $r \in R$ such that $P_r$ is correct, then \autoref{alg:instrumentation_search} will return $r'$ such that $P_{r'}$ is correct.
\end{lemma}



\section{Instrumentation Operators for Arrays}
\label{sec:instrumentation_operators_for_arrays}



\subsection{Instrumentation Operators for Quantification over Arrays}
\label{sec:normal-quantifiers}
\ja{Say that we need normalised programs}
\begin{table}[tb]
    \caption{Extension of the core language with quantified expressions.}
    \label{tab:programs_quantifiers}
    \begin{align*}
        \nonTerm{Expr} ~::=~~ &\texttt{(}\lambda \texttt{(}\nonTerm{Variable}\texttt{,}\nonTerm{Variable}\texttt{)}\texttt{.}\nonTerm{Expr}\texttt{)}~\texttt{(}\nonTerm{Expr},~\nonTerm{Expr}\texttt{)} \mid
        \\
        &\texttt{forall}\texttt{(}\nonTerm{Expr}\texttt{,} \nonTerm{Expr}\texttt{,} \nonTerm{Expr}\texttt{,} \lambda\texttt{(}\nonTerm{Variable},\nonTerm{Variable}\texttt{)}.\nonTerm{Expr}\texttt{)} \mid
        \\
        &\texttt{exists}\texttt{(}\nonTerm{Expr}\texttt{,} \nonTerm{Expr}\texttt{,} \nonTerm{Expr}\texttt{,} \lambda \texttt{(}\nonTerm{Variable},\nonTerm{Variable}\texttt{)}.\nonTerm{Expr}\texttt{)}
    \end{align*}\vspace*{-4ex}
\end{table}
To handle quantifiers in a programming setting, we extend the language defined in \autoref{tab:programs} by adding quantified expressions over arrays, as shown in \autoref{tab:programs_quantifiers}. As seen, we also extend the language with a lambda expression over two variables. The rationale for this is that a quantified property can be expressed as a binary predicate with the first argument corresponding to the value of an element and the second to the index. This allows us to express properties over both the value of an element and its index. For example, we can express that each element should be
equal to its index, as is done in the example program in \autoref{fig:quantified_example}. In the program, each element in the array is assigned the value corresponding to its index, after which it is asserted that this property indeed holds.

\pr{The dot is used inconsistently in $\lambda$ terms. How about defining a macro for lambda terms, including also the right spacing?}
Using $\texttt{P(x$_0$,i$_0$)}$ as shorthand for $\texttt{(}\mathtt{\lambda}\texttt{(x,i).P)(x$_0$,i$_0$)}$, the new expressions can be defined formally as:\pr{backslash only for forall, not exists? Ja: removed backslash again, saw now that aggregate doesnt have it.}
\begin{align*}
    \llbracket \texttt{forall}\texttt{(a, l, u, }\lambda \texttt{(x,i).P}\texttt{)} \rrbracket_s
    &~=~
    \forall \mathtt{i \in [l, u).}\; \llbracket \texttt{P(a[i],i)} \rrbracket_s
    \\
    \llbracket \texttt{exists}\texttt{(a, l, u, }\lambda \texttt{(x,i).P}\texttt{)} \rrbracket_s
    &~=~
    \exists \mathtt{i \in [l, u).}\; \llbracket \texttt{P(a[i],i)} \rrbracket_s
\end{align*}
Note that the types of \texttt{x} and \texttt{a} must be compatible and \texttt{P} a Boolean-valued expression. 

\begin{figure}[tb]
    \centering
\begin{lstlisting}[escapechar=@,style=ExtWhileStyle]
Int N = nondet;
assume(N > 0);
Array Int a = const(0, N);
Int i = 0;
while(i < N) {
    a = store(a, i, i);
    i = i + 1;
}
Bool b = forall(a, 0, N, @$\lambda$@(i,x).(x == i));
assert(b);
\end{lstlisting}
    \caption{Example of program to be verified using a quantified assert statement.}
    \label{fig:quantified_example}
\end{figure}


\begin{figure}[!htb]
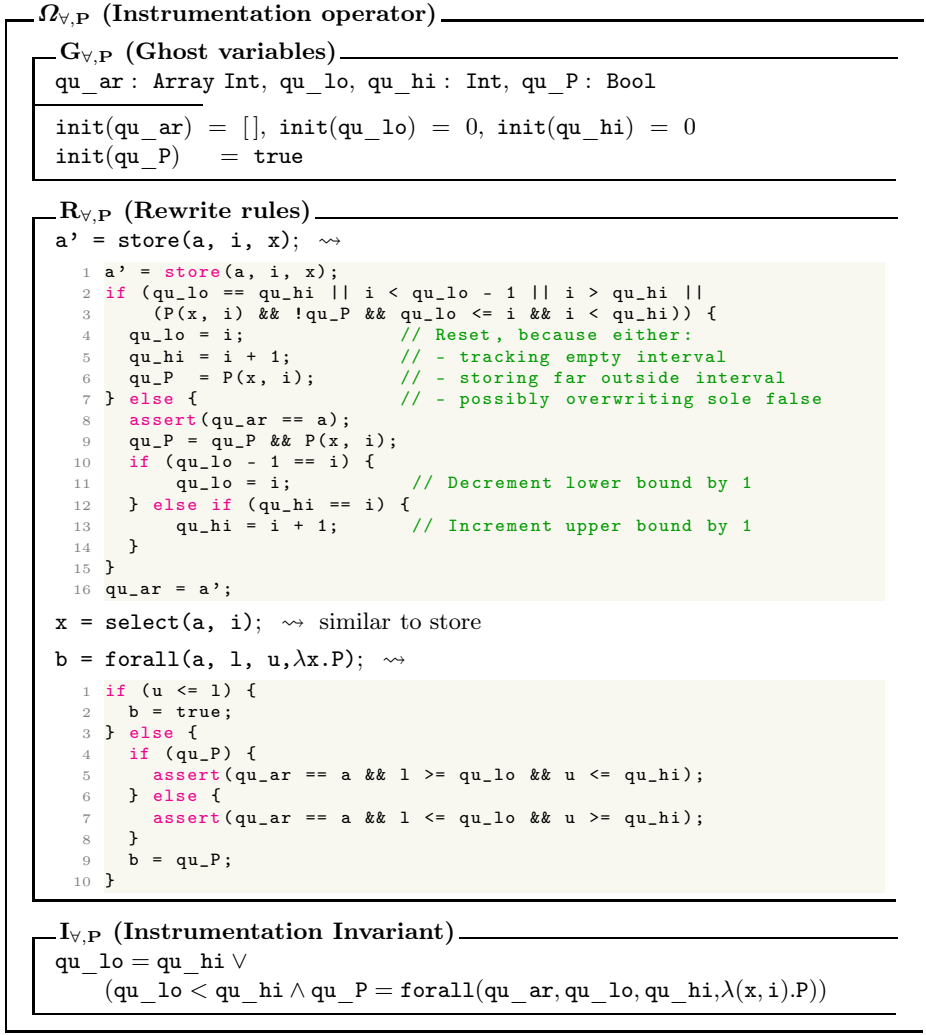

    \centering
\footnotesize
\begin{class}{\bm{\Omega}_{\mathbf{\forall,P}} \textbf{
(Instrumentation operator)}}
\begin{schema}{\mathbf{G_{\forall,P}} \textbf{ (Ghost variables)}}
    \mathtt{qu\_ar:~Array~Int},\
    \mathtt{qu\_lo,~qu\_hi:~Int},\ 
    \mathtt{qu\_P:~Bool}
    \where
    \mathtt{init(qu\_ar)} ~=~ [\,],\ 
    \mathtt{init(qu\_lo)} ~=~ 0,\ 
    \mathtt{init(qu\_hi)} ~=~ 0 \\
    \mathtt{init(qu\_P)} ~~~~=~ \mathtt{true}
\end{schema}\\
\begin{schema}{\mathbf{R_{\forall,P}} \textbf{ (Rewrite~rules)}}
     \texttt{a' = store(a, i, x)}; ~\leadsto~  \\
     \qquad \mbox{
\begin{lstlisting}[linewidth=0.85\textwidth, style=ExtWhileStyleOp]
a' = store(a, i, x);
if (qu_lo == qu_hi || i < qu_lo - 1 || i > qu_hi || 
    (P(x, i) && !qu_P && qu_lo <= i && i < qu_hi)) {
  qu_lo = i;             // Reset, because either:
  qu_hi = i + 1;         // - tracking empty interval
  qu_P  = P(x, i);       // - storing far outside interval
} else {                 // - possibly overwriting sole false
  assert(qu_ar == a);
  qu_P = qu_P && P(x, i);
  if (qu_lo - 1 == i) {
      qu_lo = i;          // Decrement lower bound by 1
  } else if (qu_hi == i) {
      qu_hi = i + 1;      // Increment upper bound by 1
  }
}
qu_ar = a';
\end{lstlisting}}
     \also
     \texttt{x = select(a, i)}; ~\leadsto~  \text{similar to store} 
     \also
     \texttt{b = forall(a, l, u,} \lambda \texttt{x.P)}; ~\leadsto~ \\
     \qquad \mbox{
\begin{lstlisting}[linewidth=0.85\textwidth, style=ExtWhileStyleOp]
if (u <= l) {
  b = true;
} else {
  if (qu_P) {
    assert(qu_ar == a && l >= qu_lo && u <= qu_hi);
  } else {
    assert(qu_ar == a && l <= qu_lo && u >= qu_hi);
  }
  b = qu_P;
}
\end{lstlisting}}
\end{schema}\\
\begin{schema}{\mathbf{I_{\forall,P}}\textbf{ (Instrumentation Invariant)}}
     \mathtt{qu\_lo} = \mathtt{qu\_hi} \: \lor \\ \qquad
     (\mathtt{qu\_lo} < \mathtt{qu\_hi}\land 
     \mathtt{qu\_P} = 
     \mathtt{forall(qu\_ar, qu\_lo, qu\_hi,} \lambda\mathtt{(x, i).P))}
\end{schema}
\end{class}
\caption{Definition of an instrumentation operator for universal quantification}
    \label{fig:my_label}\vspace*{-2ex}
\end{figure}


\ja{ToAdd: figure with rewrite rules, and figures for instrumented program}%
To handle programs such as the one in \autoref{fig:quantified_example}, we turn to the instrumentation framework outlined in \autoref{sec:instrumentation_operators},
which we use here to define an instrumentation operator for universal quantification.
The general idea is to instrument programs with a ghost variable,
tracking if some predicate holds for all elements in an interval of the array,
with shadow variables representing the tracked array, and the bounds of the interval.
Naturally, an instrumentation operator for existential quantification can be defined in a similar fashion.
For simplicity, we shall assume a \emph{normal form} of programs, into which every program can be rewritten by introducing additional variables. In the normal form, \lstinline{store}, \lstinline{select} and \lstinline{forall} can only occur in simple assignment statements. For example, stores are restricted to occur in statements of the form: \lstinline{a' = store(a, i, x)}.

Over such normalised programs, and for a universally quantified expression  $\texttt{forall(a, l, u, }\lambda\texttt{(x,i)(P))}$, we define the instrumentation operator
\instropdef{\forall, P} as shown in \autoref{fig:my_label} over four ghost variables. The array over which quantification occurs is tracked by $\mathtt{qu\_ar}$
and the variables~$\mathtt{qu\_lo}$, $\mathtt{qu\_hi}$ represent the bounds of the currently tracked interval.
The result of the quantified expression is tracked by $\mathtt{qu\_P}$, whose value is $\mathit{true}$ iff $\mathtt{P}$ holds for all elements in $\mathtt{a}$ in the interval $[\mathtt{qu\_lo, qu\_hi})$. The rewrite rules for stores, selects and assignments of universally quantified expressions are then defined
as follows.
For stores, the first if-branch resets the tracking to the one element interval $[\mathtt{i}, \mathtt{i+1})$ when accessing elements far outside of the currently tracked interval, or if we are tracking the empty interval (as is the case at initialisation).
If an access occurs immediately adjacent
to the currently tracked interval (e.g., if $\mathtt{i} = \mathtt{qu\_lo-1}$), then that element is added to the tracked interval, and the value of $\mathtt{qu\_P}$ is updated to also account for the value of $\mathtt{P}$ at index $\mathtt{i}$.
If instead the access is within the tracked interval, then we either reset the interval (if $\mathtt{qu\_P}$ is $\mathtt{false}$) or keep the interval unchanged (if $\mathtt{qu\_P}$ is $\mathtt{true}$).
Rewrites of selects are similar to stores, except tracking does not
need to be reset when reading inside the tracked interval.
For rewrites of quantified expressions, if the quantified interval is empty, $\mathtt{b}$ is assigned $\mathtt{true}$.
Otherwise, assertions check that the tracked interval matches the quantified interval before assigning $\mathtt{t}$ to $\mathtt{qu\_P}$.
If $\mathtt{qu\_P}$ is $\mathtt{true}$, then it is sufficient that quantification occurs over a sub-interval of the tracked interval, and vice versa if $\mathtt{qu\_P}$ is $\mathtt{false}$.

\pr{this paragraph should be made more positive!}
The result of applying $\Omega_{\forall, P}$ to the program in \autoref{fig:quantified_example} is shown in
\iftr Appendix~\ref{app:instrexample_quantifier}\else \cite{techreport}\fi.
As exhibited by the experiments in \autoref{sec:evaluation}, the resulting program is in many cases \dg{Isn't this too weak a statement?} easier to verify by state-of-the-art verification tools.
%
Note that the instrumentation operator defined is only one possibility among many. For example, one could track several ranges simultaneously over the array in question, or also track the index of some element in the array over which \texttt{P} holds, or make different choices on stores outside of the tracked interval.
\ja{Last sentence might be covered elsewhere in paper.
CL: I think it is only covered in the appendix, when explaining alternatives
for max, so it is good to have it here (and doesn't hurt to also mention it
again for max I guess).}

%
The following lemma establishes correctness of the instrumentation operator.
The proof can be found in
\iftr Appendix~\ref{app:correctnessForall}\else \cite{techreport}\fi.
%
%
\begin{lemma}[Correctness of $\Omega_\mathit{\forall,P}$]
\label{lem:forall-correctness}
$\Omega_\mathit{\forall,P}$ is an instrumentation operator, i.e., it
adheres to the constraints imposed in \autoref{def:instr-op}.
\end{lemma}


\subsection{Instrumentation Operators for Aggregation over Arrays}\label{sec:instrumentation-models}

We now turn to the verification of safety properties with \emph{aggregation.} As examples of aggregation, we consider in particular the operators \verb!\sum! and \verb!\max!, calculating the sum and maximum value of an array, respectively. Aggregation is supported in the form of \emph{extended quantifiers} in the specification languages JML~\cite{DBLP:books/daglib/p/LeavensBR99} and ACSL~\cite{acsl}, and is frequently needed for the specification of functional correctness properties.  Although commonly used, most verification tools do not support aggregation, so that properties involving aggregation have to be manually rewritten using standard quantifiers, pure recursive functions, or ghost code involving loops. This reduction step is error-prone, and represents an additional complication for automatic verification approaches, but can be handled elegantly using the instrumentation framework.
%
%
For generality, we formalise aggregation over arrays with the help of monoid homomorphisms.
\begin{definition}[Monoid]
  A \emph{monoid} is a structure~$(M, \circ, e)$ consisting of a non-empty set~$M$, a binary associative operation~$\circ$ on $M$, and a neutral element~$e \in M$.
  A \emph{monoid} is \emph{commutative} if $\circ$ is commutative. 
  A monoid is \emph{cancellative} if $x \circ y = x \circ z$ implies $y = z$, and $y \circ x = z \circ x$ implies $y = z$, for all $x, y, z \in M$.
\end{definition}

For aggregation, we model finite intervals of arrays using the cancellative monoid~$(D^*, \cdot, \epsilon)$ of finite sequences over some data domain~$D$. The concatenation operator~$\cdot$ is  non-commutative.

\begin{definition}[Monoid Homomorphism]
  A \emph{monoid homomorphism} is a function
  $h : M_1 \to M_2$ between monoids~$(M_1, \circ_1, e_1)$ and $(M_2, \circ_2, e_2)$ with the properties $h(x \circ_1 y) = h(x) \circ_2 h(y)$ and $h(e_1) = e_2$.
\end{definition}

Ordinary quantifiers can be modelled as homormorphisms~$D^* \to \mathbbm{B}$, so that the instrumentation in this section strictly generalizes \autoref{sec:normal-quantifiers}.
%
%
%
A second classical example is the computation of the \emph{maximum} (similarly, \emph{minimum}) value in a sequence. For the domain of integers, the natural monoid to use is the algebra~$(\mathbbm{Z}_{-\infty}, \max, -\infty)$ of integers extended with $-\infty$,\footnote{For machine integers, $-\infty$ could be replaced with \texttt{INT\_MIN}.} and the homomorphism~$h_{\max}$ is generated by mapping singleton sequences~$\langle n \rangle$ to the value~$n$.
A third example is the computation of the element \emph{sum} of an integer sequence, corresponding to the monoid~$(\mathit{Z}, +, 0)$ and the homomorphism~$h_{\operatorname{sum}}$. 
Similarly, the \emph{number of occurrences} of some element can be computed. The considered monoid in the last two cases of aggregation is even cancellative.

\vspace*{-2ex}
\subsubsection{Programming Language with Aggregation}
\label{sec:programming-language}

%
We extend our core
programming language with expressions
$\texttt{aggregate}_{M, h}\texttt{(}\nonTerm{Expr}\texttt{,} \nonTerm{Expr}\texttt{,} \nonTerm{Expr}\texttt{)}$,
and use monoid homomorphisms
to formalise them.
Recall that we denote by $D_{\sigma}$ the domain of a program type~$\sigma$.
%
%
%
\begin{definition}
  Let~$\texttt{Array}~\sigma$ be an array type, $\sigma_M$ a program type,
  $M$~a commutative monoid that is a subset of $D_{\sigma_M}$, and $h : D_{\sigma}^* \to M$ a monoid homomorphism. 
  Let furthermore $\mathit{ar}$ be an expression of type $\texttt{Array}~\sigma$, and $l$ and $u$ integer expressions. 
  Then, $\texttt{aggregate}_{M, h}\texttt{(}\mathit{ar}\texttt{,} l\texttt{,} u\texttt{)}$ is an expression of type~$\sigma_M$,
with semantics defined by:
\begin{equation*}
    \llbracket \texttt{aggregate}_{M, h}\texttt{(}\mathit{ar}\texttt{,} l\texttt{,} u\texttt{)} \rrbracket_s
    ~=~
    h(\langle \llbracket \mathit{ar} \rrbracket_s (\llbracket l \rrbracket_s), \llbracket \mathit{ar} \rrbracket_s (\llbracket l \rrbracket_s + 1), \ldots, \llbracket \mathit{ar} \rrbracket_s (\llbracket u \rrbracket_s - 1) \rangle)
\end{equation*}
\end{definition}
Intuitively, the expression~$\texttt{aggregate}_{M, h}\texttt{(}\mathit{ar}\texttt{,} l\texttt{,} u\texttt{)}$ denotes the result of applying the homomorphism~$h$ to the slice~$\mathit{ar}[l \;..\; u-1]$ of the array~$\mathit{ar}$.
As a convention, in case $u < l$ we assume that the result of \texttt{aggregate} is $h(\langle\rangle)$.
As with array accesses, we assume also that \lstinline{aggregate} only occurs
in normalised statements of the form $\texttt{t = aggregate}_{M, h}\texttt{(}\mathit{ar}\texttt{,} l\texttt{,} u\texttt{)}$.

In our examples, we use derived operations as found in ACSL:
 \verb!\max! as short-hand notation for $\texttt{aggregate}_{(\mathbbm{Z}_{-\infty}, \max, -\infty), h_{\max}}$\footnote{With a slight abuse of the framework, we assume that $\mathbbm{Z}_{-\infty}$ is represented by the program type~\texttt{Int}, mapping $-\infty$ to some fixed integer number. More elegant solutions are not difficult to devise, but add unnecessary complexity.}, and \verb!\sum! as short-hand notation for $\texttt{aggregate}_{(\mathbbm{Z}, +, 0), h_{\operatorname{sum}}}$.


\subsubsection{An Instrumentation Operator for Maximum}
\label{sec:instr-op-max}

For \verb!\max!,
an operator \instropdef{max} can be defined similarly to the operator $\Omega_{\forall, P}$
from \autoref{sec:normal-quantifiers}, in that the maximum value in a
particular interval of the array is tracked.
One key difference is that an extra ghost variable
\lstinline{ag_max_idx} is added to track an array
index where the maximum value of the array interval is stored,
in order to not have to reset tracking on every store inside
of the tracked interval.
A complete definition is proposed in
\iftr Appendix~\ref{app:correctnessMax}\else \cite{techreport}\fi.

\subsubsection{An instrumentation operator for Sum}
\label{sec:cancellative}

Cancellative aggregation is aggregation based on a
cancellative monoid.
Cancellative aggregation makes it possible to 
track aggregate values faithfully even when storing
\emph{inside} of the tracked interval, unlike \lstinline{\max}
and universal quantification.
An example of a cancellative operator is the aggregate \lstinline{\sum}.
%
%

\begin{figure}[!htb]
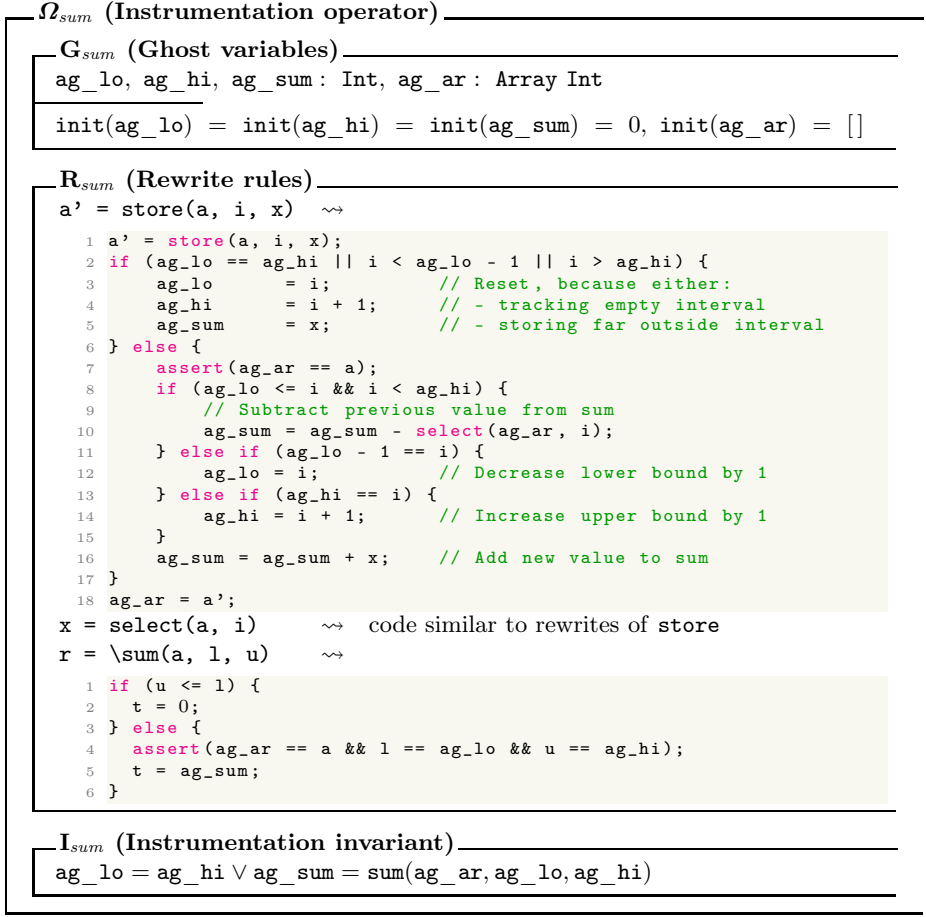

    \centering
\footnotesize
\begin{class}{\bm{\Omega}_{\mathit{sum}} \textbf{
(Instrumentation operator)}}
\begin{schema}{\mathbf{G_{\mathit{sum}}} \textbf{ (Ghost variables)}}
    \mathtt{ag\_lo},~\mathtt{ag\_hi},~\mathtt{ag\_sum:~Int},
    \ \mathtt{ag\_ar:~Array\ Int}
    \where
    \mathtt{init(ag\_lo)} ~=~
    \mathtt{init(ag\_hi)} ~=~
    \mathtt{init(ag\_sum)} ~=~ 0,\ 
    \mathtt{init(ag\_ar)} ~=~ [\,] 
\end{schema}
\\ 
\begin{schema}{\mathbf{R_{\mathit{sum}}} \textbf{ (Rewrite~rules)}}
\begin{array}{l@{\quad}l@{\quad}l@{\quad\qquad}l}
     \texttt{a' = store(a, i, x)}  &\leadsto&  & \\
     \multicolumn{4}{l}{\hspace{2em}\mbox{
\begin{lstlisting}[linewidth=0.85\textwidth, style=ExtWhileStyleOp]
a' = store(a, i, x);
if (ag_lo == ag_hi || i < ag_lo - 1 || i > ag_hi) {
    ag_lo      = i;         // Reset, because either:
    ag_hi      = i + 1;     // - tracking empty interval
    ag_sum     = x;         // - storing far outside interval
} else {
    assert(ag_ar == a);
    if (ag_lo <= i && i < ag_hi) {
        // Subtract previous value from sum
        ag_sum = ag_sum - select(ag_ar, i);
    } else if (ag_lo - 1 == i) {
        ag_lo = i;          // Decrease lower bound by 1
    } else if (ag_hi == i) {
        ag_hi = i + 1;      // Increase upper bound by 1
    }
    ag_sum = ag_sum + x;    // Add new value to sum
}
ag_ar = a';
\end{lstlisting}}}
    \\
     \texttt{x = select(a, i)}  &\leadsto&
        \text{code similar to rewrites of \texttt{store}} & 
    \\
     \texttt{r = \textbackslash sum(a, l, u)}  &\leadsto&  & \\
     \multicolumn{4}{l}{\hspace{2em}\mbox{
\begin{lstlisting}[linewidth=0.85\textwidth, style=ExtWhileStyleOp]
if (u <= l) {
  t = $0$;
} else {
  assert(ag_ar == a && l == ag_lo && u == ag_hi);
  t = ag_sum;
}
\end{lstlisting}}}\\
\end{array}
\end{schema}
\\ 
\begin{schema}{\mathbf{I_{\mathit{sum}}}\textbf{ (Instrumentation invariant)}}
     \mathtt{ag\_lo = ag\_hi} \lor \mathtt{ag\_sum = sum(ag\_ar, ag\_lo, ag\_hi)}
\end{schema}
\end{class}
\caption{Definition of an instrument operator $\Omega_{sum}$}
    \label{fig:sum-instr-op}
\end{figure}


The instrumentation operator \instropdef{sum} is defined
in \autoref{fig:sum-instr-op}.
\iftrue 
The instrumentation code tracks the sum of values in the interval,
and when increasing the bounds of the tracked interval,
the new values are simply added to the tracked sum.
Since \lstinline{\sum} is cancellative, when storing inside
of the tracked interval, the previous value at the index
being written to is first subtracted from the sum,
before adding the new value, ensuring that the correct
aggregate value is computed.
\fi 
The following correctness result is shown in
\iftr Appendix~\ref{app:correctnessSum}\else \cite{techreport}:\fi.
%
\begin{lemma}[Correctness of $\Omega_\mathit{sum}$]
\label{lem:sum-correctness}
$\Omega_\mathit{sum}$ is an instrumentation operator, i.e., it
adheres to the constraints imposed in \autoref{def:instr-op}.
\end{lemma}

\vspace*{-2ex}
\subsubsection{Deductive Verification of Instrumentation Operators}
As stated in \autoref{sec:instrumentation_operators},
instrumentation operators may be verified independently
of the programs to be instrumented.
The operators described in this paper, i.e. square, universal quantification, maximum, and sum,
have been verified in the verification tool
Frama-C~\cite{DBLP:conf/sefm/CuoqKKPSY12}.
The verified instrumentations are adaptations for the
C~language semantics and execution model.
More specifically, the adapted operators assume C native arrays,
rather than functional ones.



\section{Evaluation}
\label{sec:evaluation}

\subsection{Implementation}
To evaluate our instrumentation framework, we have implemented the instrumentation operators for quantifiers and aggregation over arrays. The implementation is done over constrained Horn clauses (CHCs), by adding the rewrite rules defined in \autoref{sec:instrumentation_operators_for_arrays} to \eldarica~\cite{eldarica}, an open-source solver for CHCs.
We also implemented the automatic application of the instrumentation operators, largely following \autoref{alg:instrumentation_search} but with a few minor changes due to the CHC setting. The CHC setting makes our implementation available to various CHC-based verification tools, for instance \jayhorn\ (Java) \cite{jayhorn-2017}, \korn\ (C)~\cite{DBLP:conf/tacas/Ernst23}, \rusthorn\ (Rust)~\cite{rusthorn}, \seahorn\ (C/LLVM)~\cite{seahorn} and \tricera\ (C)~\cite{tricera-fmcad}.


In order to evaluate our approach at the level of C programs, we extended \tricera, an open-source assertion-based model checker that translates C programs into a set of CHCs and relies on \eldarica\ as back-end solver. \tricera\ is extended to parse quantifiers and aggregation operators in its input~C programs and to encode them as part of the translation into CHCs. We call the resulting toolchain \monocera.
An artefact that includes \monocera\ and the benchmarks is available online~\cite{artifact}.

To handle complicated access patterns, for instance a program processing an array from the beginning and end at the same time, the implementation can apply multiple instrumentation operators simultaneously; the number of operators
is incremented when \autoref{alg:instrumentation_search} returns \emph{Inconclusive}.
\ja{is this mentioned prev? If so, crossref. Should perhaps go into fig 6.}



\subsection{Experiments and Comparisons}
To assess our implementation, we assembled a test suite and carried out experiments comparing \monocera\ with the state-of-the-art C~model checkers \cpachecker~2.1.1~\cite{cpachecker}, \seahorn~10.0.0~\cite{seahorn} and \tricera~0.2.
It should be noted that deductive verification frameworks, such as Dafny and Frama-C, can handle, for example, the program in \autoref{fig:quantified_example} if they are provided with a manually written loop invariant; however, since \monocera~ relies on automatic techniques for invariant inference, we only benchmark against tools using similar automatic techniques. We also excluded \veriabs~\cite{DBLP:conf/tacas/AfzalCCCDGKMUV20}, since its licence does not permit its use for scientific evaluation. 

The tools were set up, as far as possible, with equivalent configurations; for instance, to use the SMT-LIB theory of arrays~\cite{smtlib} in order to model C arrays, and a mathematical (as opposed to machine) semantics of integers. \cpachecker\ was configured to use $k$-induction~\cite{DBLP:conf/cav/0001DW15}, which was the only configuration that worked in our tests using mathematical integers. 
\seahorn\ was run using the default settings. All tests were run on a Linux machine 
with AMD Opteron 2220 SE @ 2.8 GHz and 6 GB RAM with a timeout of 300 seconds.

\paragraph{Test Suite.}
The comparison includes a set of programs calculating properties related to the quantification and aggregation properties over arrays. The benchmarks and verification results are summarised in \autoref{tab:results_monocera}. 
The benchmark suite contains programs ranging between 16 to 117 LOC and is comprised of two parts:
\begin{inparaenum}[(i)]
    \item 117~programs taken from the SV-COMP repository~\cite{sv-benchmarks-2022}, and
    \item 26~programs crafted by the authors  (\texttt{min}: 6, \texttt{max}: 8, \texttt{sum}: 9, \texttt{forall}: 3).
\end{inparaenum}

To construct the SV-COMP benchmark set for \monocera\, we gathered all test files from the directories prefixed with \lstinline$array$ or \lstinline$loop$, and singled out programs containing some  assert statement that could be rewritten using a quantifier or an aggregation operator over a single array. For example, loops
\begin{center}
\lstinline[style=ExtWhileStyleOp, basicstyle=\ttfamily\small,]!for (int i = 0; i < N; i++) assert(a[i] <= 0);! 
\end{center}
can be rewritten using $\mathtt{forall}$ or $\mathtt{max}$ operators.
We created a benchmark for each possible rewriting; for instance,
in the case of $\mathtt{max}$, by rewriting the loop into
\lstinline[basicstyle={\small\ttfamily}]!assert(\max(a, 0, N) <= 0)!.
The original benchmarks were used for the evaluation of the other tools, none of which supported (extended) quantifiers.
%
%

In (ii), we crafted 9 programs that make use of aggregation or quantifiers, and derived further benchmarks by considering different array sizes (10, 100 and unbounded size); one combination (unbounded array inside a struct) had to be excluded, as it is not valid C. In order to evaluate other tools on our crafted benchmarks, we reversed the process described for the SV-COMP benchmarks and translated the operators into corresponding loop constructs.

\iftrue
\begin{table}[tb]
\footnotesize
\setlength{\belowrulesep }{0pt}
\setlength{\aboverulesep }{0pt}
  \begin{center}
    \begin{tabular}{
    l@{\hspace{0pt}}
    c@{\hspace{6pt}}
    c
    c
    c
    cc@{\hspace{12pt}}
    c
    c
    c
    c@{\hspace{8pt}}
    c@{\hspace{2pt}}
    c
    c@{\hspace{8pt}}
    c
    c
    }
& \multicolumn{5}{c}{\textbf{Verification results}}&&
\multicolumn{3}{@{\hspace{0pt}}c}{\textbf{Ver. time}} && \multicolumn{2}{@{\hspace{-2pt}}c}{\textbf{Inst. space}} && \multicolumn{2}{@{\hspace{-4pt}}c}{\textbf{Inst. steps}}
\\ \toprule
&  {\#Tests} & \monoc & \tri & \sea & \cpa  && Min & Max & Avg && Max & Avg && Max & Avg    
\\ \cmidrule(r{4pt}){2-2}\cmidrule(l{0pt}){3-6}\cmidrule{8-10}\cmidrule(r{3pt}){12-13}\cmidrule(r{3pt}){15-16}
\texttt{min}     & 17  & 9  & 2 & 2 & 2 && 22 & 59  &  33 && 27 & 11 && 55 & 24 
\\
\texttt{max}  & 12   & 8 & 2 & 3 & 3  && 21  & 285 & 76 && 108 & 21 && 96 & 30 
\\
\texttt{sum}  & 26  & 16 & 3 & 3 & 3 && 26   & 245  & 78 && 2916 & 188 && 284 & 36 
\\
\texttt{forall}& 96 & 30 & 1 & 0 & 2 && 14  & 236 & 91 && 59049 & 2446 && 334 & 59 \\
\bottomrule
    \end{tabular}
    \end{center}
    \caption{Results for \monocera\ (\monoc), \tricera\ (\tri), \seahorn\ (\sea), and \cpachecker\ (\cpa). For \monocera, also statistics are given for verification time (s), size of the instrumentation search space, and search iterations.}
    \label{tab:results_monocera}
    \vspace*{-3ex}
\end{table}
\fi

\paragraph{Results.}
In \autoref{tab:results_monocera}, we present the number of verified programs per instrumentation operator for each tool, as well as further statistics for \monocera\ regarding verification times and instrumentation search space. The ``Inst. space'' column indicates the size of the instrumentation search space (i.e., number of instrumentations producible by applying the non-deterministic instrumentation operator). ``Inst. steps'' column indicates the number of attempted instrumentations, i.e., number of iterations in the while-loop in \autoref{alg:instrumentation_search}. In some cases there are more iterations than the size of the search space, which is because the check in \autoref{alg:instrumentation_search}, line~5 can time out, and might be attempted again at a later point in our implementation.
In 
\iftr Appendix~\ref{app:evaluation_results}\else \cite{techreport}\fi, 
we provide tables listing all results per benchmark for each tool.

For the SV-COMP benchmarks, \cpachecker\ managed to verify 1 program, while \seahorn\ and \tricera\ could not verify any programs. \monocera\ verified in total 42 programs from SV-COMP. 
Regarding the crafted benchmarks, several tools could verify the examples with array size 10. However, when the array size was 100\ja{todo: double check this} or unbounded, only \monocera~succeeded. 

\section{Related Work}

It is common practice, in both model checking and deductive verification, to translate high-level specifications to low-level specifications prior to verification (e.g., \cite{DBLP:reference/mc/2018,cell2010,DBLP:conf/sas/BjornerMR13,DBLP:conf/sas/MonniauxG16}). Such translations often make use of ghost variables and ghost code, although relatively little systematic research has been done on the required properties of ghost code~\cite{DBLP:journals/fmsd/FilliatreGP16}. The addition of ghost variables to a program for tracking the value of complex expressions also has similarities with the concept of term abstraction in Horn solving~\cite{DBLP:conf/lpar/AlbertiBGRS12}. To the best of our knowledge, we are presenting the first general framework for automatic program instrumentation.

A lot of research in \emph{software model checking} considered the handling of standard quantifiers~$\forall, \exists$ over arrays. In the setting of constrained Horn clauses, properties with universal quantifiers can sometimes be reduced to quantifier-free reasoning over non-linear Horn clauses~\cite{DBLP:conf/sas/BjornerMR13,DBLP:conf/sas/MonniauxG16}. Our approach follows the same philosophy of applying an up-front program transformation, but in a more general setting.
Various direct approaches to infer quantified array invariants have been proposed as well: e.g., by extending the IC3 algorithm~\cite{DBLP:conf/atva/GurfinkelSV18}, syntax-guided synthesis~\cite{DBLP:conf/cav/FedyukovichPMG19}, learning~\cite{DBLP:conf/cav/0001LMN13}, by solving recurrence equations~\cite{DBLP:conf/lpar/HenzingerHKR10},
backward reachability~\cite{DBLP:conf/lpar/AlbertiBGRS12}, or superposition~\cite{DBLP:conf/fmcad/GeorgiouGK20}. To the best of our knowledge, such methods have not been extended to aggregation.

\emph{Deductive verification} tools usually have rich support for quantified specifications, but rely on auxiliary assertions like loop invariants provided by the user, and on SMT solvers or automated theorem provers for quantifier reasoning.
Although several deductive verification tools can parse extended quantifiers, few offer support for reasoning about them. Our work is closest to the method for handling comprehension operators in Spec$\#$~\cite{DBLP:conf/sac/LeinoM09}, which relies on code annotations provided by the user, but provides heuristics to automatically verify such annotations. The code instrumentation presented in this paper has similarity with the proof rules in Spec$\#$; the main differences are that our method is based on an upfront program transformation, and that we aim at automatically finding required program invariants, as opposed to only verifying their correctness. The KeY tool provides proof rules similar to the ones in Spec$\#$ for some of the JML extended quantifiers~\cite{DBLP:series/lncs/10001}; those proof rules can be applied manually to verify human-written invariants. The Frama-C system~\cite{DBLP:conf/sefm/CuoqKKPSY12} can parse ACSL extended quantifiers~\cite{acsl}, but, to the best of our knowledge, none of the Frama-C plugins can automatically process such quantifiers. Other systems, e.g., Dafny~\cite{dafny}, require users to manually define aggregation operators as recursive functions.

In the theory of \emph{algebraic data-types}, several transformation-based approaches have been proposed to verify properties that involve recursive functions or catamorphisms~\cite{DBLP:journals/pacmpl/KSG22,DBLP:journals/tplp/AngelisPFP22}. Aggregation over arrays resembles the evaluation of recursive functions over data-types; a major difference is that data-types are more restricted with respect to accessing and updating data than arrays.

Array folds logic (AFL)~\cite{DBLP:conf/cav/DacaHK16} is a decidable logic in which properties on arrays beyond standard quantification can be expressed: for instance, counting the number of elements with some property. Similar properties can be expressed using automata on data words~\cite{DBLP:conf/csl/Segoufin06}, or in variants of monadic second-order logic~\cite{DBLP:journals/tocl/NevenSV04}. Such languages can be seen as alternative formalisms to aggregation or extended quantifiers; they do not cover, however, all kinds of aggregation we are interested in. Array sums cannot be expressed in AFL or data automata, for instance.

\section{Conclusion}
\label{sec:conclusion}

We have presented a framework for automatic and provably correct program instrumentation, allowing the automatic verification of programs containing certain expressive language constructs, which are not directly supported by the existing automatic verification tools. 
Our experiments with a prototypical implementation, in the tool \monocera, show that our method is able to automatically verify a significant number of benchmark programs involving quantification and aggregation over arrays that are beyond the scope of other tools.

At this point, there are still various other benchmarks that \monocera\ (as well as other tools) cannot verify. We believe that many of those benchmarks are in reach of our method, because of the generality of our approach.
Since adding ghost code is known to be a powerful
specification paradigm, more powerful instrumentation operators
can be easily formulated.
Our approach is also parameterised on the applied instrumentation
operators, meaning it can be adapted to any kind of program
by tailoring the right instrumentation operator.

In future work, we plan to develop a library of instrumentation operators for a wide range of problematic language constructs, including various classes of arithmetic expressions such as non-linear arithmetic,
other types of structures with regular access patterns
such as binary heaps,
and general linked-data structures.
We also plan to improve the evaluation by extending the benchmark suite. In a preliminary evaluation on a set of unsafe test programs from the SV-COMP library\iftr\else(which can be found in~\cite{techreport})\fi,
the results show that using the instrumentation framework does not outperform state-of-the art model checkers for such programs and further experiments are needed to identify how our framework can be used in this context.
Another line of work is the establishment of stronger completeness results than the weak completeness result presented here, for specific programming language fragments. 

\paragraph{Acknowledgements.}
This work has been partially funded by the Swedish  Vinnova FFI Programme under grant 2021-02519, the Swedish Research Council (VR)
    under grant~2018-04727, the Swedish Foundation for Strategic
    Research (SSF) under the project WebSec (Ref.\ RIT17-0011), and the
    Wallenberg project UPDATE. We are also grateful for the opportunity
    to discuss the research at the Dagstuhl Seminar~22451 on ``Principles of Contract Languages.''


\clearpage
\bibliographystyle{splncs04}
\bibliography{references}

\begin{thebibliography}{10}
\providecommand{\url}[1]{\texttt{#1}}
\providecommand{\urlprefix}{URL }
\providecommand{\doi}[1]{https://doi.org/#1}

\bibitem{DBLP:conf/tacas/AfzalCCCDGKMUV20}
Afzal, M., Chakraborty, S., Chauhan, A., Chimdyalwar, B., Darke, P., Gupta, A.,
  Kumar, S., M, C.B., Unadkat, D., Venkatesh, R.: Veriabs : Verification by
  abstraction and test generation (competition contribution). In: Biere, A.,
  Parker, D. (eds.) Tools and Algorithms for the Construction and Analysis of
  Systems - 26th International Conference, {TACAS} 2020, Held as Part of the
  European Joint Conferences on Theory and Practice of Software, {ETAPS} 2020,
  Dublin, Ireland, April 25-30, 2020, Proceedings, Part {II}. Lecture Notes in
  Computer Science, vol. 12079, pp. 383--387. Springer (2020),
  \url{https://doi.org/10.1007/978-3-030-45237-7\_25}

\bibitem{DBLP:series/lncs/10001}
Ahrendt, W., Beckert, B., Bubel, R., H{\"{a}}hnle, R., Schmitt, P.H., Ulbrich,
  M. (eds.): Deductive Software Verification - The {KeY} Book - From Theory to
  Practice, Lecture Notes in Computer Science, vol. 10001. Springer (2016),
  \url{https://doi.org/10.1007/978-3-319-49812-6}

\bibitem{DBLP:conf/lpar/AlbertiBGRS12}
Alberti, F., Bruttomesso, R., Ghilardi, S., Ranise, S., Sharygina, N.: Lazy
  abstraction with interpolants for arrays. In: Bj{\o}rner, N.S., Voronkov, A.
  (eds.) Logic for Programming, Artificial Intelligence, and Reasoning - 18th
  International Conference, LPAR-18, M{\'{e}}rida, Venezuela, March 11-15,
  2012. Proceedings. Lecture Notes in Computer Science, vol.~7180, pp. 46--61.
  Springer (2012), \url{https://doi.org/10.1007/978-3-642-28717-6\_7}

\bibitem{artifact}
Amilon, J., Esen, Z., Gurov, D., Lidström, C., Rümmer, P.: {Artifact for the
  CAV 2023 paper "Automatic Program Instrumentation for Automatic
  Verification"} (Apr 2023), \url{https://doi.org/10.5281/zenodo.7875416}

\bibitem{smtlib}
Barrett, C., Fontaine, P., Tinelli, C.: {The SMT-LIB Standard: Version 2.6}.
  Tech. rep., Department of Computer Science, The University of Iowa (2017),
  available at {\tt www.SMT-LIB.org}

\bibitem{acsl}
Baudin, P., Filli\^{a}tre, J.C., March\'{e}, C., Monate, B., Moy, Y., Prevosto,
  V.: {ACSL}: {ANSI/ISO C} Specification Language,
  \url{http://frama-c.com/acsl.html}

\bibitem{sv-report-22}
Beyer, D.: Progress on software verification: {SV-COMP} 2022. In: Fisman, D.,
  Rosu, G. (eds.) Tools and Algorithms for the Construction and Analysis of
  Systems - 28th International Conference, {TACAS} 2022, Held as Part of the
  European Joint Conferences on Theory and Practice of Software, {ETAPS} 2022,
  Munich, Germany, April 2-7, 2022, Proceedings, Part {II}. Lecture Notes in
  Computer Science, vol. 13244, pp. 375--402. Springer (2022),
  \url{https://doi.org/10.1007/978-3-030-99527-0\_20}

\bibitem{sv-benchmarks-2022}
Beyer, D.: {SV-Benchmarks: Benchmark Set for Software Verification and Testing
  ({SV-COMP} 2022 and {Test-Comp} 2022)} (Jan 2022),
  \url{https://doi.org/10.5281/zenodo.5831003}

\bibitem{DBLP:conf/cav/0001DW15}
Beyer, D., Dangl, M., Wendler, P.: Boosting k-induction with
  continuously-refined invariants. In: Kroening, D., Pasareanu, C.S. (eds.)
  Computer Aided Verification - 27th International Conference, {CAV} 2015, San
  Francisco, CA, USA, July 18-24, 2015, Proceedings, Part {I}. Lecture Notes in
  Computer Science, vol.~9206, pp. 622--640. Springer (2015),
  \url{https://doi.org/10.1007/978-3-319-21690-4\_42}

\bibitem{cpachecker}
Beyer, D., Keremoglu, M.E.: {CPAchecker}: A tool for configurable software
  verification. In: Gopalakrishnan, G., Qadeer, S. (eds.) Computer Aided
  Verification - 23rd International Conference, {CAV} 2011, Snowbird, UT, USA,
  July 14-20, 2011. Proceedings. Lecture Notes in Computer Science, vol.~6806,
  pp. 184--190. Springer (2011),
  \url{https://doi.org/10.1007/978-3-642-22110-1\_16}

\bibitem{DBLP:conf/birthday/BjornerGMR15}
Bj{\o}rner, N., Gurfinkel, A., McMillan, K.L., Rybalchenko, A.: Horn clause
  solvers for program verification. In: Beklemishev, L.D., Blass, A.,
  Dershowitz, N., Finkbeiner, B., Schulte, W. (eds.) Fields of Logic and
  Computation {II} - Essays Dedicated to Yuri Gurevich on the Occasion of His
  75th Birthday. Lecture Notes in Computer Science, vol.~9300, pp. 24--51.
  Springer (2015), \url{https://doi.org/10.1007/978-3-319-23534-9\_2}

\bibitem{DBLP:conf/sas/BjornerMR13}
Bj{\o}rner, N.S., McMillan, K.L., Rybalchenko, A.: On solving universally
  quantified {H}orn clauses. In: Logozzo, F., F{\"{a}}hndrich, M. (eds.) Static
  Analysis - 20th International Symposium, {SAS} 2013, Seattle, WA, USA, June
  20-22, 2013. Proceedings. Lecture Notes in Computer Science, vol.~7935, pp.
  105--125. Springer (2013), \url{https://doi.org/10.1007/978-3-642-38856-9\_8}

\bibitem{DBLP:reference/mc/2018}
Clarke, E.M., Henzinger, T.A., Veith, H., Bloem, R. (eds.): Handbook of Model
  Checking. Springer (2018), \url{https://doi.org/10.1007/978-3-319-10575-8}

\bibitem{DBLP:conf/sefm/CuoqKKPSY12}
Cuoq, P., Kirchner, F., Kosmatov, N., Prevosto, V., Signoles, J., Yakobowski,
  B.: {Frama-C} - {A} software analysis perspective. In: Eleftherakis, G.,
  Hinchey, M., Holcombe, M. (eds.) Software Engineering and Formal Methods -
  10th International Conference, {SEFM} 2012, Thessaloniki, Greece, October
  1-5, 2012. Proceedings. Lecture Notes in Computer Science, vol.~7504, pp.
  233--247. Springer (2012),
  \url{https://doi.org/10.1007/978-3-642-33826-7\_16}

\bibitem{DBLP:conf/cav/DacaHK16}
Daca, P., Henzinger, T.A., Kupriyanov, A.: Array folds logic. In: Chaudhuri,
  S., Farzan, A. (eds.) Computer Aided Verification - 28th International
  Conference, {CAV} 2016, Toronto, ON, Canada, July 17-23, 2016, Proceedings,
  Part {II}. Lecture Notes in Computer Science, vol.~9780, pp. 230--248.
  Springer (2016), \url{https://doi.org/10.1007/978-3-319-41540-6\_13}

\bibitem{DBLP:journals/tplp/AngelisPFP22}
{De Angelis}, E., Proietti, M., Fioravanti, F., Pettorossi, A.: Verifying
  catamorphism-based contracts using constrained {H}orn clauses. Theory Pract.
  Log. Program.  \textbf{22}(4),  555--572 (2022),
  \url{https://doi.org/10.1017/S1471068422000175}

\bibitem{cell2010}
Donaldson, A.F., Kroening, D., R{\"u}mmer, P.: Automatic analysis of
  scratch-pad memory code for heterogeneous multicore processors. In: Esparza,
  J., Majumdar, R. (eds.) Tools and Algorithms for the Construction and
  Analysis of Systems. LNCS, vol.~6015, pp. 280--295. Springer (2010)

\bibitem{DBLP:conf/tacas/Ernst23}
Ernst, G.: Korn - software verification with {H}orn clauses (competition
  contribution). In: Sankaranarayanan, S., Sharygina, N. (eds.) Tools and
  Algorithms for the Construction and Analysis of Systems - 29th International
  Conference, {TACAS} 2023, Held as Part of the European Joint Conferences on
  Theory and Practice of Software, {ETAPS} 2022, Paris, France, April 22-27,
  2023, Proceedings, Part {II}. Lecture Notes in Computer Science, vol. 13994,
  pp. 559--564. Springer (2023),
  \url{https://doi.org/10.1007/978-3-031-30820-8\_36}

\bibitem{tricera-fmcad}
Esen, Z., R{\"{u}}mmer, P.: {TriCera}: Verifying {C} programs using the theory
  of heaps. In: 2022 Formal Methods in Computer Aided Design, {FMCAD} 2022,
  Trento, Italy, October 17 - October 21, 2022 (2022), (To appear)

\bibitem{DBLP:conf/cav/FedyukovichPMG19}
Fedyukovich, G., Prabhu, S., Madhukar, K., Gupta, A.: Quantified invariants via
  syntax-guided synthesis. In: Dillig, I., Tasiran, S. (eds.) Computer Aided
  Verification - 31st International Conference, {CAV} 2019, New York City, NY,
  USA, July 15-18, 2019, Proceedings, Part {I}. Lecture Notes in Computer
  Science, vol. 11561, pp. 259--277. Springer (2019),
  \url{https://doi.org/10.1007/978-3-030-25540-4\_14}

\bibitem{DBLP:journals/fmsd/FilliatreGP16}
Filli{\^{a}}tre, J., Gondelman, L., Paskevich, A.: The spirit of ghost code.
  Formal Methods Syst. Des.  \textbf{48}(3),  152--174 (2016),
  \url{https://doi.org/10.1007/s10703-016-0243-x}

\bibitem{fla-sax-01-popl}
Flanagan, C., Saxe, J.B.: Avoiding exponential explosion: generating compact
  verification conditions. In: Hankin, C., Schmidt, D. (eds.) Proceedings of:
  Symposium on Principles of Programming Languages ({POPL'01}). pp. 193--205.
  {ACM} (2001). \doi{10.1145/360204.360220}

\bibitem{DBLP:conf/cav/0001LMN13}
Garg, P., L{\"{o}}ding, C., Madhusudan, P., Neider, D.: Learning universally
  quantified invariants of linear data structures. In: Sharygina, N., Veith, H.
  (eds.) Computer Aided Verification - 25th International Conference, {CAV}
  2013, Saint Petersburg, Russia, July 13-19, 2013. Proceedings. Lecture Notes
  in Computer Science, vol.~8044, pp. 813--829. Springer (2013),
  \url{https://doi.org/10.1007/978-3-642-39799-8\_57}

\bibitem{DBLP:conf/fmcad/GeorgiouGK20}
Georgiou, P., Gleiss, B., Kov{\'{a}}cs, L.: Trace logic for inductive loop
  reasoning. In: 2020 Formal Methods in Computer Aided Design, {FMCAD} 2020,
  Haifa, Israel, September 21-24, 2020. pp. 255--263. {IEEE} (2020),
  \url{https://doi.org/10.34727/2020/isbn.978-3-85448-042-6\_33}

\bibitem{seahorn}
Gurfinkel, A., Kahsai, T., Komuravelli, A., Navas, J.A.: {The SeaHorn
  Verification Framework}. In: Kroening, D., Pasareanu, C.S. (eds.) Computer
  Aided Verification - 27th International Conference, {CAV} 2015, San
  Francisco, CA, USA, July 18-24, 2015, Proceedings, Part {I}. Lecture Notes in
  Computer Science, vol.~9206, pp. 343--361. Springer (2015),
  \url{https://doi.org/10.1007/978-3-319-21690-4\_20}

\bibitem{DBLP:conf/atva/GurfinkelSV18}
Gurfinkel, A., Shoham, S., Vizel, Y.: Quantifiers on demand. In: Lahiri, S.K.,
  Wang, C. (eds.) Automated Technology for Verification and Analysis - 16th
  International Symposium, {ATVA} 2018, Los Angeles, CA, USA, October 7-10,
  2018, Proceedings. Lecture Notes in Computer Science, vol. 11138, pp.
  248--266. Springer (2018),
  \url{https://doi.org/10.1007/978-3-030-01090-4\_15}

\bibitem{DBLP:books/daglib/0022394}
Harrison, J.: Handbook of Practical Logic and Automated Reasoning. Cambridge
  University Press (2009)

\bibitem{DBLP:conf/lpar/HenzingerHKR10}
Henzinger, T.A., Hottelier, T., Kov{\'{a}}cs, L., Rybalchenko, A.: Aligators
  for arrays (tool paper). In: Ferm{\"{u}}ller, C.G., Voronkov, A. (eds.) Logic
  for Programming, Artificial Intelligence, and Reasoning - 17th International
  Conference, LPAR-17, Yogyakarta, Indonesia, October 10-15, 2010. Proceedings.
  Lecture Notes in Computer Science, vol.~6397, pp. 348--356. Springer (2010),
  \url{https://doi.org/10.1007/978-3-642-16242-8\_25}

\bibitem{eldarica}
Hojjat, H., R{\"u}mmer, P.: The {{ELDARICA}} {H}orn solver. In: FMCAD 2018.
  pp.~1--7 (2018). \doi{10.23919/FMCAD.2018.8603013}

\bibitem{DBLP:journals/pacmpl/KSG22}
K., H.G.V., Shoham, S., Gurfinkel, A.: Solving constrained {H}orn clauses
  modulo algebraic data types and recursive functions. Proc. {ACM} Program.
  Lang.  \textbf{6}({POPL}),  1--29 (2022),
  \url{https://doi.org/10.1145/3498722}

\bibitem{jayhorn-2017}
Kahsai, T., Kersten, R., R{\"{u}}mmer, P., Sch{\"{a}}f, M.: Quantified heap
  invariants for object-oriented programs. In: Eiter, T., Sands, D. (eds.)
  LPAR-21, 21st International Conference on Logic for Programming, Artificial
  Intelligence and Reasoning, Maun, Botswana, May 7-12, 2017. EPiC Series in
  Computing, vol.~46, pp. 368--384. EasyChair (2017),
  \url{https://easychair.org/publications/paper/Pmh}

\bibitem{DBLP:books/daglib/p/LeavensBR99}
Leavens, G.T., Baker, A.L., Ruby, C.: {JML:} {A} notation for detailed design.
  In: Kilov, H., Rumpe, B., Simmonds, I. (eds.) Behavioral Specifications of
  Businesses and Systems, The Kluwer International Series in Engineering and
  Computer Science, vol.~523, pp. 175--188. Springer (1999),
  \url{https://doi.org/10.1007/978-1-4615-5229-1\_12}

\bibitem{dafny}
Leino, K.R.M.: Dafny: An automatic program verifier for functional correctness.
  In: Clarke, E.M., Voronkov, A. (eds.) Logic for Programming, Artificial
  Intelligence, and Reasoning - 16th International Conference, LPAR-16, Dakar,
  Senegal, April 25-May 1, 2010, Revised Selected Papers. Lecture Notes in
  Computer Science, vol.~6355, pp. 348--370. Springer (2010),
  \url{https://doi.org/10.1007/978-3-642-17511-4\_20}

\bibitem{DBLP:conf/sac/LeinoM09}
Leino, K.R.M., Monahan, R.: Reasoning about comprehensions with first-order
  {SMT} solvers. In: Shin, S.Y., Ossowski, S. (eds.) Proceedings of the 2009
  {ACM} Symposium on Applied Computing (SAC), Honolulu, Hawaii, USA, March
  9-12, 2009. pp. 615--622. {ACM} (2009),
  \url{https://doi.org/10.1145/1529282.1529411}

\bibitem{rusthorn}
Matsushita, Y., Tsukada, T., Kobayashi, N.: {RustHorn}: {CHC}-based
  verification for {Rust} programs. {ACM} Trans. Program. Lang. Syst.
  \textbf{43}(4),  15:1--15:54 (2021), \url{https://doi.org/10.1145/3462205}

\bibitem{DBLP:conf/sas/MonniauxG16}
Monniaux, D., Gonnord, L.: Cell morphing: From array programs to array-free
  {H}orn clauses. In: Rival, X. (ed.) Static Analysis - 23rd International
  Symposium, {SAS} 2016, Edinburgh, UK, September 8-10, 2016, Proceedings.
  Lecture Notes in Computer Science, vol.~9837, pp. 361--382. Springer (2016),
  \url{https://doi.org/10.1007/978-3-662-53413-7\_18}

\bibitem{DBLP:journals/tocl/NevenSV04}
Neven, F., Schwentick, T., Vianu, V.: Finite state machines for strings over
  infinite alphabets. {ACM} Trans. Comput. Log.  \textbf{5}(3),  403--435
  (2004), \url{https://doi.org/10.1145/1013560.1013562}

\bibitem{DBLP:conf/atva/PriyaZSVBG21}
Priya, S., Zhou, X., Su, Y., Vizel, Y., Bao, Y., Gurfinkel, A.: Verifying
  verified code. In: Hou, Z., Ganesh, V. (eds.) Automated Technology for
  Verification and Analysis - 19th International Symposium, {ATVA} 2021, Gold
  Coast, QLD, Australia, October 18-22, 2021, Proceedings. Lecture Notes in
  Computer Science, vol. 12971, pp. 187--202. Springer (2021),
  \url{https://doi.org/10.1007/978-3-030-88885-5\_13}

\bibitem{DBLP:conf/lics/Reynolds02}
Reynolds, J.C.: Separation logic: {A} logic for shared mutable data structures.
  In: 17th {IEEE} Symposium on Logic in Computer Science {(LICS} 2002), 22-25
  July 2002, Copenhagen, Denmark, Proceedings. pp. 55--74. {IEEE} Computer
  Society (2002), \url{https://doi.org/10.1109/LICS.2002.1029817}

\bibitem{DBLP:conf/csl/Segoufin06}
Segoufin, L.: Automata and logics for words and trees over an infinite
  alphabet. In: {\'{E}}sik, Z. (ed.) Computer Science Logic, 20th International
  Workshop, {CSL} 2006, 15th Annual Conference of the EACSL, Szeged, Hungary,
  September 25-29, 2006, Proceedings. Lecture Notes in Computer Science,
  vol.~4207, pp. 41--57. Springer (2006),
  \url{https://doi.org/10.1007/11874683\_3}

\end{thebibliography}

\iftr

\newpage
\appendix

\section{Typing Rules of the Core Language}
\label{app:typing-rules}

The typing rules of the core language are presented in~\autoref{tab:typing}.

\begin{table}[h]
    \caption{Typing rules of the core language.}
    \label{tab:typing}
    
    \begin{gather*}
      \infer{\typeJudg{\nonTerm{DecimalNumber}}{\texttt{Int}}}{}
      \quad
      \infer{\typeJudg{\texttt{true}}{\texttt{Bool}}}{}
      \quad
      \infer{\typeJudg{\texttt{false}}{\texttt{Bool}}}{}
      \quad
      \infer{\vphantom{X}\typeJudg{x}{\sigma}}{x \in \mathcal{X}, \alpha(x) = \sigma}
      \\[1ex]
      \infer{\typeJudg{s \eqeq t}{\texttt{Bool}}}{\typeJudg{s}{\sigma} & \typeJudg{t}{\sigma}}
      \quad
      \infer{\typeJudg{s  \;\texttt{<=}\; t}{\texttt{Bool}}}{\typeJudg{s}{\texttt{Int}} & \typeJudg{t}{\texttt{Int}}}
      \quad
      \infer{\typeJudg{s  \;\texttt{+}\; t}{\texttt{Int}}}{\typeJudg{s}{\texttt{Int}} & \typeJudg{t}{\texttt{Int}}}
      \quad
      \infer{\typeJudg{s  \;\texttt{*}\; t}{\texttt{Int}}}{\typeJudg{s}{\texttt{Int}} & \typeJudg{t}{\texttt{Int}}}
      \\[1ex]
     \infer{\typeJudg{ \texttt{!} c}{\texttt{Bool}}}{\typeJudg{c}{\texttt{Bool}}}
      \quad
      \infer{\typeJudg{s \;\texttt{\&\&}\; t}{\texttt{Bool}}}{\typeJudg{s}{\texttt{Bool}} & \typeJudg{t}{\texttt{Bool}}}
      \quad
      \infer{\typeJudg{s \;\texttt{||}\; t}{\texttt{Bool}}}{\typeJudg{s}{\texttt{Bool}} & \typeJudg{t}{\texttt{Bool}}}
      \\[1ex]
      \infer{\typeJudg{\texttt{select(}a\texttt{,} t\texttt{)}}{\sigma}}{\typeJudg{a}{\texttt{Array}~\sigma} & \typeJudg{t}{\texttt{Int}}}
      \quad
      \infer{\typeJudg{\texttt{store(}a\texttt{,} s\texttt{,} t\texttt{)}}{\texttt{Array}~\sigma}}{\typeJudg{a}{\texttt{Array}~\sigma} & \typeJudg{s}{\texttt{Int}} & \typeJudg{t}{\sigma}}
    \end{gather*}
\end{table}

\section{Correctness of Instrumentation Search}
\label{app:instrSearchCorrectness}

\paragraph{Proof of \autoref{lem:searchCorrectness}}

\autoref{alg:instrumentation_search} will return an instrumentation~$r$
of when it has derived that $P_r$ is correct; due to the soundness
of instrumentation operators, then also $P$ is correct.

\autoref{alg:instrumentation_search} will return $\mathit{Incorrect}$ only
when it has discovered a counterexample for $P_r$ that ends in a failing assertion
that also occurs in $P$, i.e., that has not been introduced as a part of
instrumentation. Due to the weak completeness of instrumentation operators,
then also $P$ is incorrect.

Assuming that there is an $r$ such that $P_r$ is correct, the correctness
also of $P$ follows. \autoref{alg:instrumentation_search} can then not
return the result $\mathit{Incorrect}$. To see that \autoref{alg:instrumentation_search}
will eventually find some instrumentation~$r'$ such that $P_{r'}$ is correct, note
that the algorithm will in lines~12--13 only eliminate instrumentations~$r''$
such that $P_{r''}$ is incorrect.



\section{Example of Instrumented Program for Universal Quantification}
\label{app:instrexample_quantifier}
An example of an instrumentation using the instrumentation operator $\Sigma_{\forall,P}$ is presented in \autoref{fig:quantified_example_instr}.
\begin{figure}[tb]
    \centering
\begin{lstlisting}[escapechar=@,style=ExtWhileStyle]
Int qu_lo = 0; qu_hi = 0;
Int qu_ar = [];
Bool qu_P = true;
Int N = nondet;
assume(N > 0);
Array Int a = const(0, N);
Int i = 0;
while(i < N) {
    a' = store(a, i, x);
    if (qu_lo == qu_hi || i < qu_lo - 1 || i > qu_hi || 
        (P(x, i) && !qu_P && qu_lo <= i && i < qu_hi)) {
      qu_lo = i;         // Reset, because either:
      qu_hi = i + 1;     // - tracking empty interval
      qu_P  = P(x, i);   // - storing far outside interval
    } else {             // - possibly overwriting sole false
      assert(qu_ar == a);
      qu_P = qu_P && P(x, i);
      if (qu_lo - 1 == i) {
          qu_lo = i;          // Decrement lower bound by 1
      } else if (qu_hi == i) {
          qu_hi = i + 1;      // Increment upper bound by 1
      }
    }
    qu_ar = a';
    a = store(a, i, i);
    i = i + 1;
}
Bool b;
if (u <= l) {
  b = true;
} else {
  if (qu_P) {
    assert(qu_ar == a $\wedge$ l >= qu_lo $\wedge$ u <= qu_hi);
  } else {
    assert(qu_ar == a $\wedge$ l <= qu_lo $\wedge$ u >= qu_hi);
  }
  b = qu_P;
}
assert(b);
\end{lstlisting}
    \caption{Resulting program from applying the instrumentation $\Omega_{\forall,P}$ to the program in \autoref{fig:quantified_example}.}
    \label{fig:quantified_example_instr}
\end{figure}

\clearpage

\section{Correctness of Instrumentation Operator for 
    Universal Quantification}
\label{app:correctnessForall}

\paragraph{Proof (\autoref{lem:forall-correctness}).}

First, observe that \lstinline{qu_lo} and \lstinline{qu_hi}
are initialised to the same value, so the instrumentation invariant
is established.
It then remains to show that the rewrite rules adhere
to the remaining constraints. The case where we rewrite normalised
\lstinline{select} statements is left out, as it is similar to
\lstinline{store}.

It is clear that all the instrumented code terminates, and that
rewrites of \lstinline{store} only assign to ghost variables
and to \lstinline{a'},
and rewrites of \lstinline{forall} expressions only
to the assigned variable~\lstinline{b}.

Rewrites of \lstinline{forall} do not assign to ghost variables,
and so preserve~$I$.
For rewrites of \lstinline{store}, we assume $I$, and for the proof
we treat the assertion in the rewrite rule as an assumption.
In the cases of an empty interval, a store outside of the tracked
interval, and when
possibly overwriting a sole \lstinline{false} value,
since we reset the bounds,
and the tracked Boolean value, to only track a single element,
then \lstinline{qu_P} has the correct value in the resulting array
with the new bounds,
and since we assign \lstinline{qu_ar} the result of the store,
$I$~is preserved.
In the remaining cases, by the invariant and the assertions
\lstinline{qu_P} is assigned the correct truth value of the array
being stored to, within the given bounds, by conjunction of the
previous quantified truth value and the evaluation of the predicate
on the new element.
If the store occurs just outside the tracked bounds,
also the bounds are updated, and $I$ is preserved.

Finally, in rewrites of \lstinline{store}, the assignment
\lstinline{a' = store(a, i, x);} is not changed, but only additional
code is added that does not assign to program variables,
and so the semantics is preserved.
In rewrites of \lstinline{forall}, we assume $I$ and the assertion
in the rewrite rule.
Since by the invariant \lstinline{qu_P} is the correct value
before the store, over
\lstinline{qu_ar} in the tracked interval, and since by the assertion
\lstinline{qu_ar} is the same array as in the quantified
expression, then \lstinline{qu_P} is also
the correct truth value in \lstinline{a} within the tracked bounds.
Then, if tracked value is \lstinline{true} for the tracked interval,
it must also be \lstinline{true} for smaller intervals, and if
it is \lstinline{false} it is also \lstinline{false} for large
intervals, and we assert this relation between the tracked bounds
and the actual bounds of the quantified expression.
Thus, \lstinline{b} is assigned
the result of the quantified expression, and the semantics is preserved.
\hfill \qed

\section{An Instrumentation Operator for 
    \texttt{\textbackslash max}}
\label{app:correctnessMax}


\begin{figure}[!htb]
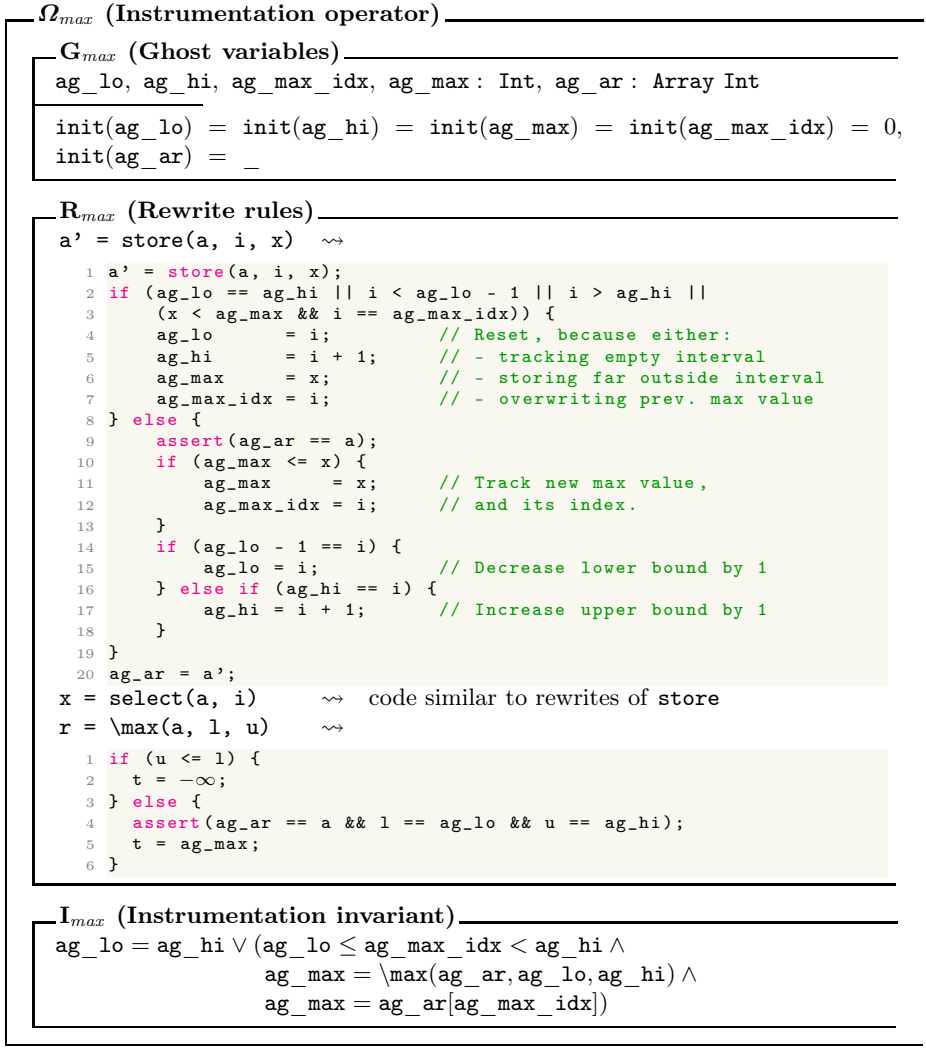

    \centering
\footnotesize
\begin{class}{\bm{\Omega}_{\mathit{max}} \textbf{
(Instrumentation operator)}}
\begin{schema}{\mathbf{G_{\mathit{max}}} \textbf{ (Ghost variables)}}
    \mathtt{ag\_lo},~\mathtt{ag\_hi},~\mathtt{ag\_max\_idx},~
    \mathtt{ag\_max:~Int},
    \ \mathtt{ag\_ar:~Array\ Int}
    \where
    \mathtt{init(ag\_lo)} ~=~
    \mathtt{init(ag\_hi)} ~=~
    \mathtt{init(ag\_max)} ~=~
    \mathtt{init(ag\_max\_idx)} ~=~ 0,\\
    \mathtt{init(ag\_ar)} ~=~ \_ 
\end{schema}
\\ 
\begin{schema}{\mathbf{R_{\mathit{max}}} \textbf{ (Rewrite~rules)}}
\begin{array}{l@{\quad}l@{\quad}l@{\quad\qquad}l}
     \texttt{a' = store(a, i, x)}  &\leadsto&  & \\
     \multicolumn{4}{l}{\hspace{2em}\mbox{
\begin{lstlisting}[linewidth=0.85\textwidth, style=ExtWhileStyleOp]
a' = store(a, i, x);
if (ag_lo == ag_hi || i < ag_lo - 1 || i > ag_hi ||
    (x < ag_max && i == ag_max_idx)) {
    ag_lo      = i;         // Reset, because either:
    ag_hi      = i + 1;     // - tracking empty interval
    ag_max     = x;         // - storing far outside interval
    ag_max_idx = i;         // - overwriting prev. max value
} else {
    assert(ag_ar == a);
    if (ag_max <= x) {
        ag_max     = x;     // Track new max value,
        ag_max_idx = i;     // and its index.
    }
    if (ag_lo - 1 == i) {
        ag_lo = i;          // Decrease lower bound by 1
    } else if (ag_hi == i) {
        ag_hi = i + 1;      // Increase upper bound by 1
    }
}
ag_ar = a';
\end{lstlisting}}}\\
     \texttt{x = select(a, i)}  &\leadsto&
        \text{code similar to rewrites of \texttt{store}} & \\
     \texttt{r = \textbackslash max(a, l, u)}  &\leadsto&  & \\
     \multicolumn{4}{l}{\hspace{2em}\mbox{
\begin{lstlisting}[linewidth=0.85\textwidth, style=ExtWhileStyleOp]
if (u <= l) {
  t = $-\infty$;
} else {
  assert(ag_ar == a && l == ag_lo && u == ag_hi);
  t = ag_max;
}
\end{lstlisting}}}\\
\end{array}
\end{schema}
\\ 
\begin{schema}{\mathbf{I_{\mathit{max}}}\textbf{ (Instrumentation invariant)}}
     \mathtt{ag\_lo} = \mathtt{ag\_hi} \vee 
     (\mathtt{ag\_lo} \leq \mathtt{ag\_max\_idx} < \mathtt{ag\_hi}
     \: \wedge \\
            \qquad\qquad\qquad\qquad\;\;
           \mathtt{ag\_max} = \mathtt{\backslash max}(\mathtt{ag\_ar}, \mathtt{ag\_lo}, \mathtt{ag\_hi}) \: \wedge \\
           \qquad\qquad\qquad\qquad\;\;
          \mathtt{ag\_max} = \mathtt{ag\_ar[ag\_max\_idx]})
\end{schema}
\end{class}
\vspace{-2em}
\caption{Definition of an instrument operator $\Omega_{max}$}
    \label{fig:max-instr-op}
\end{figure}


The instrumentation operator \instropdef{max} is defined
in \autoref{fig:max-instr-op}.
We assume here that the language has the special value
$-\infty$ or a similar construct denoting the least possible value.\footnotemark[2]
Initialisation of a variable to \texttt{\_} means it can
be set to an arbitrary value.
The rewrite rule for \texttt{select} statements is similar
to rewrites of \texttt{store} statements, except that the
line 1 would be the original \texttt{select} statement,
and the reset (lines 4--7) is not necessary when reading
inside of the tracked interval.

The idea behind~$\Omega_\mathit{max}$ is to keep
track of the aggregate value
in some interval of the array as it is read from and written to.
Uses of the aggregate function can be replaced with this
tracked value, asserting that the array and the bounds
of the interval of the array are equal to the array and bounds
of the aggregate.


In the instrumentation code for \lstinline{store}, we have two
main cases. In one case, we reset the tracking of the max value,
by setting bounds to only track the most recent value.
This is done because either the tracked interval is empty,
and we are not yet tracking any value, or an access is happening
that is not adjacent to the currently tracked interval,
or the previously tracked max value is overwritten.

In the other case, there is a store in an already tracked interval,
or just adjacent to it.
If that is the case, we assert that the tracked array coincides with
the one being accessed.
If the stored value is larger than the previous max value,
then we update the variables accordingly,
and if the access is adjacent to the previously tracked interval,
we increase the interval to include it.


Various design choices have been made in the
instrumentation operator. For example, when writing far outside of the
interval, instrumentation could also choose to ignore this access,
and keep tracking the old interval.
Another alternative is to assert that such writes do not occur, i.e.,
treating it as a failed instrumentation, and try another one.
There could also be several rewrite rules matching this case,
with a non-deterministic choice of which to use (if any),
and all choices could then be tried in a search for the
most fitting instrumentation (see \autoref{sec:transformation-strategies}).
Assertions added in the instrumentation code could often be replaced by conditionals, for instance ignoring an instrumented store in case the array or bounds do not match. And instead of tracking a single array interval, also more complicated subsets of an array could be tracked. All those choice are possible, and can be analysed for correctness, within the framework introduced in \autoref{sec:program-transformation}.



\begin{lemma}[Correctness of $\Omega_\mathit{max}$]
\label{lem:max-correctness}
$\Omega_\mathit{max}$ is an instrumentation operator, i.e., it
adheres to the constraints imposed in \autoref{def:instr-op}.
\end{lemma}

\paragraph{Proof.}
To start with, observe that \lstinline{ag_lo} and \lstinline{ag_hi}
are initialised to the same value, so the instrumentation invariant
is established.
Next, we will show that the rewrite rules adhere
to the remaining constraints. The case where we rewrite normalised
\lstinline{select} statements is left out, as it is similar to
\lstinline{store}.

It is clear that all the instrumented code terminates, and that
rewrites of \lstinline{store} only assign to ghost variables
and to \lstinline{a'},
and rewrites of \lstinline{\max} only to the assigned
variable~\lstinline{r}.

Rewrites of \lstinline{\max} do not assign to ghost variables,
and so preserve~$I_{\mathit{max}}$.
For rewrites of \lstinline{store}, we assume~$I_{\mathit{max}}$,
and for the proof
we treat the assertion in rewrite rule for stores
as an assumption.
In the cases of an empty interval, a store outside of the tracked
interval, and when
overwriting the previous max value, since we reset the bounds,
the max value, and the index to only track this element,
then \lstinline{ag_max} is the max value in the resulting array
with the new bounds,
and since we assign \lstinline{ag_ar} the result of the store,
$I_{\mathit{max}}$~is preserved.
In the remaining cases, by the invariant and the assertions
\lstinline{ag_max} is assigned the max value of the array being
stored to, within the given bounds.
If the new value is larger, we update \lstinline{ag_max}
and \lstinline{ag_max_idx} accordingly, and if the store occurs just
outside the tracked bounds, also the bounds are updated.
Then, \lstinline{ag_max} must now be the max value
after the store, within the new bounds,
and $I_{\mathit{max}}$ is preserved.

Finally, in rewrites of \lstinline{store}, the assignment
\lstinline{a' = store(a, i, x);} is not changed, but only additional
code is added that does not assign to program variables,
and so the semantics is preserved.
In rewrites of \lstinline{\max}, we assume $I_{\mathit{max}}$ and the assertion
in the rewrite rule. 
Since by the invariant \lstinline{ag_max} is the max value in
\lstinline{ag_ar} in the tracked interval, and since by the assertion
\lstinline{ag_ar} is the same array as in the aggregate
expression, and the bounds for the tracked interval are the same
as in the aggregate expression, then \lstinline{ag_max} is also
the max value in \lstinline{a} within those bounds
(and \lstinline{ag_max_idx} is a position where the value occurs).
Thus, \lstinline{r} is assigned
the result of the aggregate function, and the semantics is preserved.
\hfill \qed

\section{Correctness of Instrumentation Operator for Sum}
\label{app:correctnessSum}

Here, we show the correctness of $\Omega_{\mathit{sum}}$.

\paragraph{Proof (\autoref{lem:sum-correctness}).}
Firstly, the ghost variables \lstinline{ag_lo} and \lstinline{ag_hi} are
initialised to the same value, so $I_{\mathit{sum}}$ is established.
Now, it remains to be shown that the rewrite rules adheres
to the additional constraints.
Rewrites of normalised \lstinline{select} statements are similar
to those of \lstinline{store}, so that case is left out.

Clearly, all the instrumented code terminates, and rewrites of
\lstinline{store} only assign to ghost variables and $a'$,
and rewrites of \lstinline{\sum} only to the assigned
variable~\lstinline{r}.
Thus, all rewrites fulfill the first two conditions.

Rewrites of \lstinline{\sum} do not assign to ghost variables,
so they preserve $I_{\mathit{sum}}$.
For rewrites of \lstinline{store}, we assume $I_{\mathit{sum}}$
and for the proof treat the assertion in the rewrite rules as
an assumption.
In the cases of an empty interval, or when storing far outside
the tracked interval, since we reset the bounds, 
the sum, and the index to only track the single stored element,
then \lstinline{ag_sum} is the sum of the resulting array
with the new bounds,
and since we assign \lstinline{ag_ar} the result of the store,
$I_{\mathit{sum}}$~is preserved.
In the remaining cases, by the invariant and the assertions
\lstinline{ag_sum} is assigned the sum of the array being
stored to, within the given bounds.
When storing inside the tracked interval, we update \lstinline{ag_sum}
by first subtracting the previous value stored at that index,
and then adding the new value.
When storing adjacent to the tracked interval, we only add the
new value to \lstinline{ag_sum}, and update the bounds accordingly.
Then, \lstinline{ag_sum} must now be the sum
after the store, within the new bounds,
and $I_{\mathit{max}}$ is preserved.

Finally, in rewrites of \lstinline{store}, the assignment
\lstinline{a' = store(a, i, x);} is not changed, but only additional
code is added that does not assign to program variables,
and so the semantics is preserved.
In rewrites of \lstinline{\sum}, we assume $I_{\mathit{sum}}$ and the assertion in the rewrite rule.
Since by the invariant \lstinline{ag_sum} is the sum of
\lstinline{ag_ar} in the tracked interval, and since by the assertion
\lstinline{ag_ar} is the same array as in the aggregate
expression, and the bounds for the tracked interval are the same
as in the aggregate expression, then \lstinline{ag_sum} is also
the sum of \lstinline{a} within those bounds.
In the special case when tracking an empty interval, we immediately
assign the value $0$, which is the neutral element in the
corresponding monoid.
Thus, \lstinline{r} is assigned
the result of the aggregate function, and the semantics is preserved.
\hfill \qed

\clearpage


\section{Detailed Evaluation Results}
\label{app:evaluation_results}
We list all evaluated benchmarks in \autoref{tbl:per-benchmark-results-forall} (\emph{forall}), \autoref{tbl:per-benchmark-results-max} (\emph{max}), \autoref{tbl:per-benchmark-results-min} (\emph{min}) and \autoref{tbl:per-benchmark-results-sum} (\emph{sum}) for all evaluated tools: \monocera\ (\monoc), \tricera\ (\tri), \seahorn\ (\sea), and \cpachecker\ (\cpa). The list includes mostly SV-COMP benchmarks and the first column indicates their original directories (should be prepended with ``\texttt{c/}''). There are some crafted benchmarks that are not from SV-COMP; for these we only state the name of the benchmark (for instance \texttt{forall1-10}).

The expected result for all benchmarks in the tables is \textbf{\textcolor{green!50!black}{True}}. Durations are given next to the returned result for benchmarks which did not time out. 

On top of quantifying over arrays, many of the benchmarks have assertions over nonlinear integer arithmetic (NIA). Decision procedures for NIA are incomplete, and we have seen errors in the presence of such assertions in some of the evaluated tools.

The result ``Unknown'' is returned by \monocera\ when the instrumentation search space is exhausted.

{
\scriptsize
\renewcommand{\arraystretch}{.5}

  \begin{longtable}{lrrrr}
  \caption{\emph{forall} benchmark results for all tools. Timeout (T/O)
  is 300.0 s.}
   \label{tbl:per-benchmark-results-forall}\\
      & \cpa & \sea & \tri & \monoc\\\toprule
 \endfirsthead
      & \cpa & \sea & \tri & \monoc\\\toprule
 \endhead
     array-cav19/array\_doub\_access\_init\_const.c & T/O & T/O & T/O & \textbf{\textcolor{green!50!black}{True}} (111) \\\midrule
		array-cav19/array\_init\_nondet\_vars.c & T/O & T/O & T/O & \textbf{\textcolor{green!50!black}{True}} (14) \\\midrule
		array-cav19/array\_init\_pair\_sum\_const.c & T/O & T/O & T/O & T/O \\\midrule
		array-cav19/array\_init\_pair\_symmetr.c & T/O & T/O & T/O & T/O \\\midrule
		array-cav19/array\_init\_pair\_symmetr2.c & T/O & T/O & T/O & T/O \\\midrule
		array-cav19/array\_init\_var\_plus\_ind.c & T/O & T/O & Error (2) & \textbf{\textcolor{green!50!black}{True}} (23) \\\midrule
		array-cav19/array\_init\_var\_plus\_ind2.c & T/O & T/O & Error (2) & Unknown (109) \\\midrule
		array-cav19/array\_init\_var\_plus\_ind3.c & T/O & T/O & Error (3) & \textbf{\textcolor{green!50!black}{True}} (60) \\\midrule
		array-industry-pattern/array\_ptr\_partial\_init.c & T/O & T/O & T/O & T/O \\\midrule
		array-industry-pattern/array\_shadowinit.c & T/O & T/O & T/O & \textbf{\textcolor{green!50!black}{True}} (19) \\\midrule
		array-cav19/array\_tiling\_poly6.c & T/O & Error (0) & T/O & T/O \\\midrule
		array-cav19/array\_tripl\_access\_init\_const.c & T/O & T/O & T/O & \textbf{\textcolor{green!50!black}{True}} (199) \\\midrule
		array-fpi/condg.c & T/O & T/O & T/O & T/O \\\midrule
		array-fpi/condm.c & T/O & T/O & T/O & Error (8) \\\midrule
		array-fpi/condn.c & T/O & T/O & T/O & \textbf{\textcolor{green!50!black}{True}} (88) \\\midrule
		array-fpi/eqn1.c & T/O & Error (0) & T/O & T/O \\\midrule
		array-fpi/eqn2.c & T/O & Error (0) & T/O & T/O \\\midrule
		array-fpi/eqn3.c & T/O & Error (0) & T/O & T/O \\\midrule
		array-fpi/eqn4.c & T/O & Error (0) & T/O & T/O \\\midrule
		array-fpi/eqn5.c & T/O & Error (0) & T/O & T/O \\\midrule
		forall1-10 & \textbf{\textcolor{green!50!black}{True}} (23) & Error (0) & \textbf{\textcolor{green!50!black}{True}} (55) & \textbf{\textcolor{green!50!black}{True}} (37) \\\midrule
		forall1-100 & T/O & Error (2) & T/O & \textbf{\textcolor{green!50!black}{True}} (31) \\\midrule
		forall1-UB & T/O & Error (0) & T/O & \textbf{\textcolor{green!50!black}{True}} (32) \\\midrule
		array-fpi/ifcomp.c & T/O & Error (0) & T/O & T/O \\\midrule
		array-fpi/ifeqn1.c & T/O & Error (0) & T/O & T/O \\\midrule
		array-fpi/ifeqn2.c & T/O & Error (0) & T/O & T/O \\\midrule
		array-fpi/ifeqn3.c & T/O & Error (0) & T/O & T/O \\\midrule
		array-fpi/ifeqn4.c & T/O & Error (0) & T/O & T/O \\\midrule
		array-fpi/ifeqn5.c & T/O & Error (0) & T/O & T/O \\\midrule
		array-fpi/ifncomp.c & Error (20) & Error (0) & T/O & T/O \\\midrule
		array-fpi/indp1.c & T/O & Error (0) & T/O & T/O \\\midrule
		array-fpi/indp2.c & T/O & Error (0) & T/O & T/O \\\midrule
		array-fpi/indp3.c & T/O & Error (0) & T/O & T/O \\\midrule
		array-fpi/indp4.c & T/O & Error (0) & T/O & T/O \\\midrule
		array-tiling/mbpr2.c & T/O & T/O & T/O & T/O \\\midrule
		array-tiling/mbpr3.c & T/O & T/O & T/O & T/O \\\midrule
		array-tiling/mbpr4.c & T/O & T/O & T/O & T/O \\\midrule
		array-tiling/mbpr5.c & T/O & T/O & T/O & T/O \\\midrule
		loop-lit/mcmillan2006.c & \textbf{\textcolor{green!50!black}{True}} (7) & T/O & T/O & \textbf{\textcolor{green!50!black}{True}} (19) \\\midrule
		array-tiling/mlceu2.c & T/O & T/O & Error (3) & Error (4) \\\midrule
		array-fpi/modn.c & Error (20) & Error (0) & T/O & T/O \\\midrule
		array-fpi/modp.c & Error (17) & Error (0) & T/O & Error (9) \\\midrule
		array-fpi/mods.c & T/O & Error (0) & T/O & T/O \\\midrule
		array-fpi/ncomp.c & T/O & Error (0) & T/O & T/O \\\midrule
		array-tiling/nr2.c & T/O & T/O & T/O & \textbf{\textcolor{green!50!black}{True}} (93) \\\midrule
		array-tiling/nr3.c & T/O & T/O & T/O & \textbf{\textcolor{green!50!black}{True}} (123) \\\midrule
		array-tiling/nr4.c & T/O & T/O & T/O & \textbf{\textcolor{green!50!black}{True}} (156) \\\midrule
		array-tiling/nr5.c & T/O & T/O & T/O & \textbf{\textcolor{green!50!black}{True}} (223) \\\midrule
		array-fpi/nsqm.c & T/O & Error (0) & T/O & T/O \\\midrule
		array-fpi/nsqm-if.c & Error (15) & Error (0) & T/O & T/O \\\midrule
		array-lopstr16/partial\_lesser\_bound.c & T/O (108) & T/O & T/O & T/O \\\midrule
		array-lopstr16/partial\_lesser\_bound-1.c & T/O & Error (1) & Error (2) & \textbf{\textcolor{green!50!black}{True}} (89) \\\midrule
		array-fpi/pcomp.c & T/O & Error (0) & T/O & T/O \\\midrule
		array-tiling/pnr2.c & T/O & T/O & T/O & \textbf{\textcolor{green!50!black}{True}} (227) \\\midrule
		array-tiling/pnr3.c & T/O & T/O & T/O & \textbf{\textcolor{green!50!black}{True}} (236) \\\midrule
		array-tiling/pnr4.c & T/O & T/O & T/O & T/O \\\midrule
		array-tiling/pnr5.c & T/O & T/O & T/O & T/O \\\midrule
		array-tiling/poly1.c & T/O & Error (0) & T/O & \textbf{\textcolor{green!50!black}{True}} (25) \\\midrule
		array-tiling/poly2.c & T/O & Error (0) & T/O & T/O \\\midrule
		array-tiling/pr2.c & T/O & Error (0) & T/O & \textbf{\textcolor{green!50!black}{True}} (117) \\\midrule
		array-tiling/pr3.c & T/O & Error (0) & T/O & \textbf{\textcolor{green!50!black}{True}} (143) \\\midrule
		array-tiling/pr4.c & T/O & Error (0) & T/O & T/O \\\midrule
		array-tiling/pr5.c & T/O & Error (0) & T/O & T/O \\\midrule
		array-tiling/rew.c & T/O & T/O & T/O & T/O \\\midrule
		array-tiling/rewnif.c & T/O & T/O & T/O & \textbf{\textcolor{green!50!black}{True}} (55) \\\midrule
		array-tiling/rewnifrev.c & T/O & T/O & T/O & \textbf{\textcolor{green!50!black}{True}} (60) \\\midrule
		array-tiling/rewnifrev2.c & T/O & T/O & T/O & \textbf{\textcolor{green!50!black}{True}} (97) \\\midrule
		array-examples/sanfoundry\_27\_ground.c & T/O & T/O & T/O & Error (19) \\\midrule
		loop-crafted/simple\_array\_index\_value\_2.c & T/O & Error (1) & Error (2) & \textbf{\textcolor{green!50!black}{True}} (39) \\\midrule
		loop-crafted/simple\_array\_index\_value\_3.c & T/O & Error (1) & Error (3) & \textbf{\textcolor{green!50!black}{True}} (33) \\\midrule
		array-fpi/sina1.c & T/O & T/O & T/O & \textbf{\textcolor{green!50!black}{True}} (199) \\\midrule
		array-fpi/sina2.c & T/O & T/O & T/O & T/O \\\midrule
		array-fpi/sina3.c & T/O & T/O & T/O & T/O \\\midrule
		array-fpi/sina4.c & T/O & T/O & T/O & T/O \\\midrule
		array-fpi/sina5.c & T/O & T/O & T/O & T/O \\\midrule
		array-tiling/skipped.c & T/O & T/O & T/O & T/O \\\midrule
		array-fpi/sqm.c & T/O & Error (0) & T/O & T/O \\\midrule
		array-fpi/sqm-if.c & T/O & Error (0) & T/O & T/O \\\midrule
		array-fpi/ss2.c & Error (65) & Error (0) & T/O & T/O \\\midrule
		array-fpi/ssina.c & Error (38) & Error (0) & T/O & T/O \\\midrule
		array-examples/standard\_copyInitSum2\_ground-2.c & T/O & T/O & T/O & T/O \\\midrule
		array-examples/standard\_copyInitSum3\_ground.c & T/O & T/O & T/O & T/O \\\midrule
		array-examples/standard\_copyInit\_ground.c & T/O & T/O & T/O & T/O \\\midrule
		array-examples/standard\_init1\_ground-2.c & T/O & T/O & T/O & \textbf{\textcolor{green!50!black}{True}} (32) \\\midrule
		array-examples/standard\_init2\_ground-2.c & T/O & T/O & T/O & \textbf{\textcolor{green!50!black}{True}} (82) \\\midrule
		array-examples/standard\_init3\_ground-2.c & T/O & T/O & T/O & T/O \\\midrule
		array-examples/standard\_init4\_ground-2.c & T/O & T/O & T/O & T/O \\\midrule
		array-examples/standard\_init5\_ground-1.c & T/O & T/O & T/O & T/O \\\midrule
		array-examples/standard\_init6\_ground-2.c & T/O & T/O & T/O & T/O \\\midrule
		array-examples/standard\_init7\_ground-2.c & T/O & T/O & T/O & T/O \\\midrule
		array-examples/standard\_init8\_ground-2.c & T/O & T/O & T/O & T/O \\\midrule
		array-examples/standard\_init9\_ground-2.c & T/O & T/O & T/O & T/O \\\midrule
		array-examples/standard\_maxInArray\_ground.c & T/O & T/O & T/O & T/O \\\midrule
		array-examples/standard\_minInArray\_ground-2.c & T/O & T/O & T/O & T/O \\\midrule
		array-examples/standard\_partition\_ground-2.c & T/O & T/O & T/O & T/O \\\midrule
		array-examples/standard\_vararg\_ground.c & T/O & T/O & T/O & \textbf{\textcolor{green!50!black}{True}} (57)\\\bottomrule
  \end{longtable}

  \begin{longtable}{lrrrr}
  \caption{\emph{max} benchmark results for all tools. Timeout (T/O)
  is 300.0 s.}
   \label{tbl:per-benchmark-results-max}\\
      & \cpa & \sea & \tri & \monoc\\\toprule
 \endfirsthead
      & \cpa & \sea & \tri & \monoc\\\toprule
 \endhead
     array-cav19/array\_init\_var\_plus\_ind3.c & T/O & T/O & Error (2) & Unknown (277) \\\midrule
		battery\_diag-10 & \textbf{\textcolor{green!50!black}{True}} (155) & \textbf{\textcolor{green!50!black}{True}} (0) & T/O & T/O \\\midrule
		battery\_diag-100 & T/O & T/O & T/O & T/O \\\midrule
		array-fpi/condn.c & T/O & T/O & T/O & \textbf{\textcolor{green!50!black}{True}} (129) \\\midrule
		max\_eq-10 & \textbf{\textcolor{green!50!black}{True}} (21) & \textbf{\textcolor{green!50!black}{True}} (0) & \textbf{\textcolor{green!50!black}{True}} (78) & \textbf{\textcolor{green!50!black}{True}} (21) \\\midrule
		max\_eq-100 & T/O & T/O & T/O & \textbf{\textcolor{green!50!black}{True}} (38) \\\midrule
		max\_eq-UB & T/O & T/O & T/O & \textbf{\textcolor{green!50!black}{True}} (27) \\\midrule
		max\_leq-10 & \textbf{\textcolor{green!50!black}{True}} (25) & \textbf{\textcolor{green!50!black}{True}} (0) & \textbf{\textcolor{green!50!black}{True}} (90) & \textbf{\textcolor{green!50!black}{True}} (23) \\\midrule
		max\_leq-100 & T/O & T/O & T/O & \textbf{\textcolor{green!50!black}{True}} (30) \\\midrule
		max\_leq-UB & T/O & T/O & T/O & \textbf{\textcolor{green!50!black}{True}} (52) \\\midrule
		array-examples/sanfoundry\_27\_ground.c & T/O & T/O & T/O & \textbf{\textcolor{green!50!black}{True}} (285) \\\midrule
		array-examples/standard\_maxInArray\_ground.c & T/O & T/O & T/O & T/O\\\bottomrule
  \end{longtable}

  \begin{longtable}{lrrrr}
  \caption{\emph{min} benchmark results for all tools. Timeout (T/O)
  is 300.0 s.}
   \label{tbl:per-benchmark-results-min}\\
      & \cpa & \sea & \tri & \monoc\\\toprule
 \endfirsthead
      & \cpa & \sea & \tri & \monoc\\\toprule
 \endhead
     array-cav19/array\_doub\_access\_init\_const.c & T/O & T/O & T/O & T/O \\\midrule
		array-cav19/array\_init\_pair\_sum\_const.c & T/O & T/O & T/O & T/O \\\midrule
		array-cav19/array\_init\_pair\_symmetr2.c & T/O & T/O & T/O & T/O \\\midrule
		array-cav19/array\_init\_var\_plus\_ind.c & T/O & T/O & Error (3) & Unknown (67) \\\midrule
		array-cav19/array\_init\_var\_plus\_ind2.c & T/O & T/O & Error (2) & Unknown (64) \\\midrule
		array-cav19/array\_tripl\_access\_init\_const.c & T/O & T/O & T/O & T/O \\\midrule
		min\_eq-10 & \textbf{\textcolor{green!50!black}{True}} (25) & \textbf{\textcolor{green!50!black}{True}} (0) & \textbf{\textcolor{green!50!black}{True}} (61) & \textbf{\textcolor{green!50!black}{True}} (28) \\\midrule
		min\_eq-100 & T/O & T/O & T/O & \textbf{\textcolor{green!50!black}{True}} (22) \\\midrule
		min\_eq-UB & T/O & T/O & T/O & \textbf{\textcolor{green!50!black}{True}} (25) \\\midrule
		min\_geq-10 & \textbf{\textcolor{green!50!black}{True}} (26) & \textbf{\textcolor{green!50!black}{True}} (0) & \textbf{\textcolor{green!50!black}{True}} (51) & \textbf{\textcolor{green!50!black}{True}} (26) \\\midrule
		min\_geq-100 & T/O & T/O & T/O & \textbf{\textcolor{green!50!black}{True}} (28) \\\midrule
		min\_geq-UB & T/O & T/O & T/O & \textbf{\textcolor{green!50!black}{True}} (24) \\\midrule
		array-tiling/rew.c & T/O & T/O & T/O & \textbf{\textcolor{green!50!black}{True}} (40) \\\midrule
		array-tiling/rewnifrev.c & T/O & T/O & T/O & \textbf{\textcolor{green!50!black}{True}} (41) \\\midrule
		array-examples/standard\_minInArray\_ground-2.c & T/O & T/O & T/O & T/O \\\midrule
		array-examples/standard\_partition\_ground-2.c & T/O & T/O & T/O & T/O \\\midrule
		array-examples/standard\_vararg\_ground.c & T/O & T/O & T/O & \textbf{\textcolor{green!50!black}{True}} (59)\\\bottomrule
  \end{longtable}

  \begin{longtable}{lrrrr}
  \caption{\emph{sum} benchmark results for all tools. Timeout (T/O)
  is 300.0 s.}
   \label{tbl:per-benchmark-results-sum}\\
      & \cpa & \sea & \tri & \monoc\\\toprule
 \endfirsthead
      & \cpa & \sea & \tri & \monoc\\\toprule
 \endhead
     array-fpi/brs1.c & T/O & T/O & T/O & \textbf{\textcolor{green!50!black}{True}} (26) \\\midrule
		array-fpi/brs2.c & T/O & Error (0) & T/O & \textbf{\textcolor{green!50!black}{True}} (34) \\\midrule
		array-fpi/brs3.c & T/O & T/O & T/O & \textbf{\textcolor{green!50!black}{True}} (41) \\\midrule
		array-fpi/brs4.c & T/O & T/O & T/O & \textbf{\textcolor{green!50!black}{True}} (38) \\\midrule
		array-fpi/brs5.c & T/O & T/O & T/O & \textbf{\textcolor{green!50!black}{True}} (58) \\\midrule
		array-fpi/conda.c & T/O & T/O & T/O & T/O \\\midrule
		array-fpi/indp5.c & Error (17) & Error (0) & T/O & T/O \\\midrule
		array-fpi/ms1.c & T/O & T/O & T/O & \textbf{\textcolor{green!50!black}{True}} (26) \\\midrule
		array-fpi/ms2.c & T/O & T/O & T/O & \textbf{\textcolor{green!50!black}{True}} (46) \\\midrule
		array-fpi/ms3.c & T/O & T/O & T/O & \textbf{\textcolor{green!50!black}{True}} (31) \\\midrule
		array-fpi/ms4.c & T/O & T/O & T/O & \textbf{\textcolor{green!50!black}{True}} (33) \\\midrule
		array-fpi/ms5.c & T/O & T/O & T/O & \textbf{\textcolor{green!50!black}{True}} (32) \\\midrule
		array-fpi/s1lif.c & T/O & T/O & T/O & T/O \\\midrule
		array-fpi/s2lif.c & T/O & T/O & T/O & T/O \\\midrule
		array-fpi/s3lif.c & T/O & T/O & T/O & T/O \\\midrule
		array-fpi/s4lif.c & T/O & T/O & T/O & T/O \\\midrule
		array-fpi/s5lif.c & T/O & T/O & T/O & T/O \\\midrule
		sum\_eq-10 & \textbf{\textcolor{green!50!black}{True}} (50) & \textbf{\textcolor{green!50!black}{True}} (0) & \textbf{\textcolor{green!50!black}{True}} (45) & \textbf{\textcolor{green!50!black}{True}} (245) \\\midrule
		sum\_eq-100 & T/O & T/O & T/O & T/O \\\midrule
		sum\_eq-UB & T/O & T/O & T/O & \textbf{\textcolor{green!50!black}{True}} (55) \\\midrule
		sum\_geq-10 & \textbf{\textcolor{green!50!black}{True}} (45) & \textbf{\textcolor{green!50!black}{True}} (0) & \textbf{\textcolor{green!50!black}{True}} (98) & \textbf{\textcolor{green!50!black}{True}} (200) \\\midrule
		sum\_geq-100 & T/O & T/O & T/O & \textbf{\textcolor{green!50!black}{True}} (221) \\\midrule
		sum\_geq-UB & T/O & T/O & T/O & \textbf{\textcolor{green!50!black}{True}} (59) \\\midrule
		two\_arrays\_sum-10 & \textbf{\textcolor{green!50!black}{True}} (26) & \textbf{\textcolor{green!50!black}{True}} (0) & \textbf{\textcolor{green!50!black}{True}} (46) & T/O \\\midrule
		two\_arrays\_sum-100 & T/O & T/O & T/O & T/O \\\midrule
		two\_arrays\_sum-UB & T/O & T/O & T/O & \textbf{\textcolor{green!50!black}{True}} (106)\\\bottomrule
  \end{longtable}

}

\fi

\end{document}